\documentclass[aps, prd, superscriptaddress, preprintnumbers, showpacs]{revtex4}

\usepackage{graphicx,amsfonts,amsmath,amssymb,amstext}
\usepackage{float,wrapfig}
\usepackage{subfigure, psfrag}
\usepackage{dsfont}

\usepackage{color}
\usepackage{bm}

\newcommand{\be}{\begin{equation}}
\newcommand{\ee}{\end{equation}}
\newcommand{\ba}{\begin{eqnarray}}
\newcommand{\ea}{\end{eqnarray}}
\newcommand{\sign}{\,\mbox{sign}}

\definecolor{purple}{rgb}{0.8,0,0.6}
\definecolor{darkgreen}{rgb}{0.00,0.50,0.00}

\begin{document}

\title{Origin of dissipative Fermi arc transport in Weyl semimetals}
\date{June 11, 2016}

\author{E.~V.~Gorbar}
\affiliation{Department of Physics, Taras Shevchenko National Kiev University, Kiev, 03680, Ukraine}
\affiliation{Bogolyubov Institute for Theoretical Physics, Kiev, 03680, Ukraine}

\author{V.~A.~Miransky}
\affiliation{Department of Applied Mathematics, Western University, London, Ontario N6A 5B7, Canada}
\affiliation{Department of Physics and Astronomy, Western University, London, Ontario N6A 5B7, Canada}

\author{I.~A.~Shovkovy}
\affiliation{College of Letters and Sciences, Arizona State University, Mesa, Arizona 85212, USA}

\author{P.~O.~Sukhachov}
\affiliation{Department of Physics, Taras Shevchenko National Kiev University, Kiev, 03680, Ukraine}

\begin{abstract}
By making use of a low-energy effective model of Weyl semimetals, we show that the Fermi
arc transport is dissipative. The origin of the dissipation is the scattering of the surface Fermi arc
states into the bulk of the semimetal. It is noticeable that the corresponding scattering rate is nonzero
and can be estimated even in a perturbative theory, although in general the reliable calculations
of transport properties necessitate a nonperturbative approach. Nondecoupling of the surface
and bulk sectors in the low-energy theory of Weyl semimetals invalidates the usual argument
of a nondissipative transport due to one-dimensional arc states. This property of Weyl semimetals
is in drastic contrast to that of topological insulators, where the decoupling is protected by a gap
in the bulk. Within the framework of the linear response theory, we obtain an approximate result
for the conductivity due to the Fermi arc states and analyze its dependence on chemical potential,
temperature, and other parameters of the model.
\end{abstract}


\maketitle

\section{Introduction}
\label{sec:Introduction}

Dirac and Weyl semimetals, whose low-energy quasiparticle excitations are described by the Dirac
and Weyl equations, respectively, became an active and vibrant area of research in condensed
matter physics. Theoretically, Dirac semimetals were predicted to be realized in $\mathrm{A_3Bi}$
($\mathrm{A=Na,K,Rb}$) and $\mathrm{Cd_3As_2}$ compounds \cite{Weng,Wang}. By using
angle-resolved photoemission spectroscopy (ARPES), the Dirac nature of quasiparticles in the corresponding
materials was indeed confirmed in Refs.~\cite{Borisenko,Neupane,Liu}, opening the path toward
experimental investigations of the three-dimensional Dirac semimetals.

In momentum space, a Dirac point can be viewed as a superposition of two Weyl nodes of opposite
chirality. If these nodes are separated in momentum or energy, then a Weyl semimetal is realized.
Theoretically, the existence of Weyl semimetals was first predicted in pyrochlore iridates
in Ref.~\cite{Savrasov}. Most recently, Weyl semimetals were experimentally discovered in Refs.~\cite{Tong,Bian,Qian,Long,Belopolski,Cava}, where the observations
of characteristic surface Fermi arc states and unusual transport properties were reported for such
compounds as $\mathrm{TaAs}$, $\mathrm{TaP}$, $\mathrm{NbAs}$, $\mathrm{NbP}$, $\mathrm{Mo_xW_{1-x}Te}$, and $\mathrm{YbMnBi_2}$.

One of the key features of both Dirac and Weyl semimetals is their unusual magnetotransport, which
is profoundly affected by the chiral anomaly. The corresponding anomaly was originally discovered in
high-energy physics in the analysis of the linearly divergent triangle diagrams \cite{ABJ}. It appears
to have a qualitative effect also on magnetotransport of chiral plasmas \cite{Kharzeev}.
As was first shown by Nielsen and Ninomiya \cite{Nielsen}, the longitudinal (with respect to the
direction of the external magnetic field) magnetoresistivity in Weyl semimetals decreases with the growth of the magnetic
field (called a negative megnetoresistivity phenomenon in the literature). This phenomenon was first observed
experimentally in $\mathrm{Bi_{1-x}Sb_x}$ alloy with $\mathrm{x} \approx 0.03$ \cite{Kim:2013dia}.

The physical reason for the negative magnetoresistivity is quite transparent \cite{Nielsen,Son}. Since the
lowest Landau level (LLL) states around each Weyl node are described by an effective one-dimensional chiral theory,
they cannot backscatter on impurities. Therefore, the conductivity of such states is finite only due to
backscattering between Weyl nodes with opposite chirality. Then, since the LLL density of
states grows linearly with a magnetic field, the conductivity increases too. When the LLL contribution
dominates, the total magnetoresistivity decreases with the field. We note, however, that
theoretical analysis based on the Kubo's formula suggests that Dirac semimetals may also have a
negative magnetoresistivity in the same regime \cite{Gorbar:2013dha}. Moreover, as argued in
Ref.~\cite{Goswami:2015uxa}, the ionic impurity scattering can cause negative longitudinal
magnetoresistivity in a generic three-dimensional metal in the quantum limit.

In fact, the absence of backscattering for one-dimensional chiral fermions provides the physical reason
for the nondissipative electric transport in the quantum Hall effect (QHE) \cite{Klitzing}. It was shown 
in Ref.~\cite{Kohmoto} that the fantastic exactness of
the QHE conductivity is connected with nontrivial topological properties of QHE materials,
which are encoded in the nonzero Chern numbers. Since the vacuum has trivial topological
characteristics, current-carrying edge states are topologically protected that leads to
the celebrated bulk-boundary correspondence.

The Haldane model \cite{Haldane-model} showed that the presence of an external magnetic
field is not a necessary ingredient for the realization of the QHE. In fact, the first experimentally
discovered two-dimensional topological insulator $\mathrm{HgTe}$ \cite{Molenkamp} can be considered as
a combination of two copies of the Haldane model. Its nontrivial topological properties are
described by the $\mathbb{Z}_2$ index \cite{Kane} and its conducting edge states are composed
of two counterpropagating one-dimensional chiral fermions, related to each other through the time reversal
transformation. In the absence of magnetic impurities, the edge states of a two-dimensional topological insulator
are immune to backscattering and, thus, lead to a nondissipative transport.

A Weyl node also possesses nontrivial topological properties. It represents a monopole of the
Berry curvature in momentum space, whose charge is connected with the chirality of the node.
This is reflected in the characteristic properties of Weyl semimetals. One of them is the existence
of the surface Fermi arc states, which are open segments of the Fermi surface connecting
projections of the bulk cones onto the surface \cite{Savrasov,Aji,Haldane}. Such states connect
the Fermi surfaces for the opposite chirality quasiparticles that otherwise would appear to be
disconnected \cite{Haldane}. Experimentally, the Fermi arcs are directly observed with the help of the
ARPES technique \cite{Bian,Qian}.

Recently, the surface Fermi arcs were studied via another powerful technique,
scanning tunneling spectroscopy (STS) \cite{STS:Feenstra,STS:Crommie}.
Theoretical calculations \cite{QPI:Hasan,QPI:Kourtis,QPI:Mitchell} and experimental
studies \cite{QPI:Hasan-2,QPI:Batabyal,QPI:Inoue} of quasiparticle interference (QPI)
patterns caused by surface impurities or defects provide an alternative way to investigate
surface Fermi arcs. Note that these QPI patterns are momentum space counterparts
of Friedel oscillations \cite{QPI:Hosur}. Furthermore, the QPI reveals possible surface
scattering processes, which cannot be directly measured by ARPES. However, the
interpretation of QPI patterns strongly relays on theoretical calculations, providing a
powerful, although indirect, evidence of Fermi arcs.

It is interesting that Dirac semimetals $\mathrm{A_3Bi}$ ($\mathrm{A=Na,K,Rb}$) also have
Fermi arcs. It was shown in Ref.~\cite{Gorbar:2014sja} that there exists a hidden ``up-down'' ($ud$)
parity symmetry in the low-energy effective Hamiltonian that allows one to split the electron
states into two separate sectors, with each describing a Weyl semimetal. Therefore, the
corresponding Dirac semimetals are, in fact, $\mathbb{Z}_2$ Weyl semimetals.

As suggested in Ref.~\cite{Vishwanath}, the surface Fermi arc states can be detected indirectly by measuring 
the oscillations of the density of states in surface-sensitive probes. Indeed, when a magnetic field is applied
to a Weyl semimetal, a special type of closed magnetic orbits, involving both the surface Fermi arcs
and the bulk states, are predicted to exist \cite{Vishwanath}. The corresponding orbits produce periodic
quantum oscillations of the density of states with a characteristic dependence on the separation
between the Weyl nodes in momentum space (transport experiments that probe the electronic
properties of these orbits were suggested in Ref.~\cite{Baum}). The corresponding quantum
oscillations were recently experimentally observed \cite{Moll} in the Dirac semimetal Cd$_3$As$_2$.
As was shown in Ref.~\cite{Gorbar:2014qta}, these quantum oscillations also encode the effects
of interaction, which are predicted to modify the length of the Fermi arcs. In particular, one expects
to see a nontrivial dependence of the quantum oscillations on the orientation of the magnetic field
projection in the plane of the semimetal surface \cite{Gorbar:2014qta}. By numerically studying
a minimal model of a Weyl semimetal, it was recently shown in Ref.~\cite{Bulmash} that the area
enclosed by the Fermi surface and quantum oscillations in the thin film limit strongly depend
on the in-plane magnetic field component parallel to the Weyl nodes splitting even in the
non-interacting free theory.

Since the Fermi arcs in Weyl semimetals are topologically protected, it is natural that they are
described by an effective Hamiltonian of one-dimensional chiral fermions. This suggests that the electron
transport via Fermi arcs should be nondissipative. The latter would be a rather rigorous
conclusion if the effective one-dimensional Hamiltonian of arc states stayed generally intact in the presence
of impurities and disorder. However, as will be argued in this work, the situation is more
complicated, and this is not necessarily the case. Indeed, as was shown in Ref.~\cite{QPI:Derry},
in materials without a gap, surface quasiparticles are dephased by coupling to the
bulk. It was argued in Ref.~\cite{QPI:Mitchell} that in such a case there is no simple
effective theory for surface states because the surface Green's functions for a Weyl semimetal
necessarily contain the information about the electronic propagation from the surface into the bulk and
back to the surface.

The paper is organized as follows. The effective Hamiltonian of the surface Fermi arc states
in the simplest model of Weyl semimetal and a model of disorder used in the present paper
are described in Sec.~\ref{sec:effective-Hamiltonian}. The Fermi arc conductivity in the
effective one-dimensional surface model is considered both in naive and nonperturbative 
approaches in Sec.~\ref{sec:LinResp-Arc-surface}. Perturbative signs of dissipation in the 
full model are revealed and analyzed in Sec.~\ref{sec:Perturbative}. The quasiparticle width 
and Fermi arc conductivity in the full model are calculated in Sec.~\ref{sec:Arc-Width-Conductivity-Full}. 
The obtained results are discussed and summarized in Sec.~\ref{sec:Conclusion}.

For convenience, throughout the paper, we set $\hbar=c=1$. Here $c$ is the speed of light.

\section{Model of a Weyl semimetal and Fermi arc states}

\label{sec:effective-Hamiltonian}

In order to present our arguments as simple as possible, in this paper we study one of the simplest
continuum models of a Weyl semimetal proposed by Okugawa and Murakami \cite{Murakami}. The
Hamiltonian of the model describes a pair of Weyl nodes separated in momentum space
\begin{equation}
\label{Hamiltonian-effective-M1}
H(\mathbf{k})=\gamma\left(k_z^2 - m\right)\sigma_z+v_F\left(k_x\sigma_x+k_y\sigma_y\right),
\end{equation}
where $v_F$, $m$ and $\gamma$ are positive constants. As is easy to check, the energy 
spectrum of bulk states is given by $E(k)=\pm\sqrt{\gamma^2(k_z^2-m)^2+v_F^2(k_x^2+k_y^2)}$
and the two Weyl nodes are located at $\mathbf{k}=(0,0,\pm \sqrt{m})$. 

We assume that, in coordinate space, the semimetal is in the upper
half-plane $y>0$ and the vacuum is in the lower half-plane $y<0$. In order to prevent quasiparticles
from escaping into the vacuum region, we set $m\to -\tilde{m}$ and take the limit $\tilde{m}\to \infty$ for
$y<0$. Because of the semimetal surface at $y=0$, the translation invariance in the $y$ direction is
broken. Therefore, we keep the operator form for $k_y=-i\partial_y$ in Hamiltonian
(\ref{Hamiltonian-effective-M1}).

By considering the eigenvalue problem $H\psi=E\psi$, it is not
difficult to find the wave-functions of the surface states in semimetal and vacuum,
\begin{eqnarray}
\psi_{y>0} &=& \sqrt{p(k_z)}\, e^{ik_xx+ik_zz-p(k_z)y}\left(
                                                              \begin{array}{c}
                                                                1 \\
                                                                1 \\
                                                              \end{array}
                                                            \right),
\label{psi-surface-states}
                                                            \\
\psi_{y<0} &=& \sqrt{p(k_z)}\, e^{ik_xx+ik_zz+\gamma \tilde{m}y}\left(
                                                              \begin{array}{c}
                                                                1 \\
                                                                1 \\
                                                              \end{array}
                                                            \right),
\label{psi-All-M1}
\end{eqnarray}
respectively, where $p(k_z)= \gamma (m-k_z^2)/v_F$ and $-\sqrt{m}<k_z<\sqrt{m}$.
The corresponding expression for the surface state energy reads
\begin{equation}
E_s= v_F k_x .
\label{energy-surface}
\end{equation}
The effective Hamiltonian for the surface Fermi arc states is given by (see Ref.~\cite{Shen} for a derivation of the effective
Hamiltonian for the surface states in topological insulators)
\begin{equation}
H_{\rm surf}=-iv_F\partial_x,
\label{Hamiltonian-effective-surf-M1}
\end{equation}
and, as expected, describes one-dimensional chiral fermions.

\begin{figure*}[!ht]
\begin{center}
\includegraphics[width=0.4\textwidth]{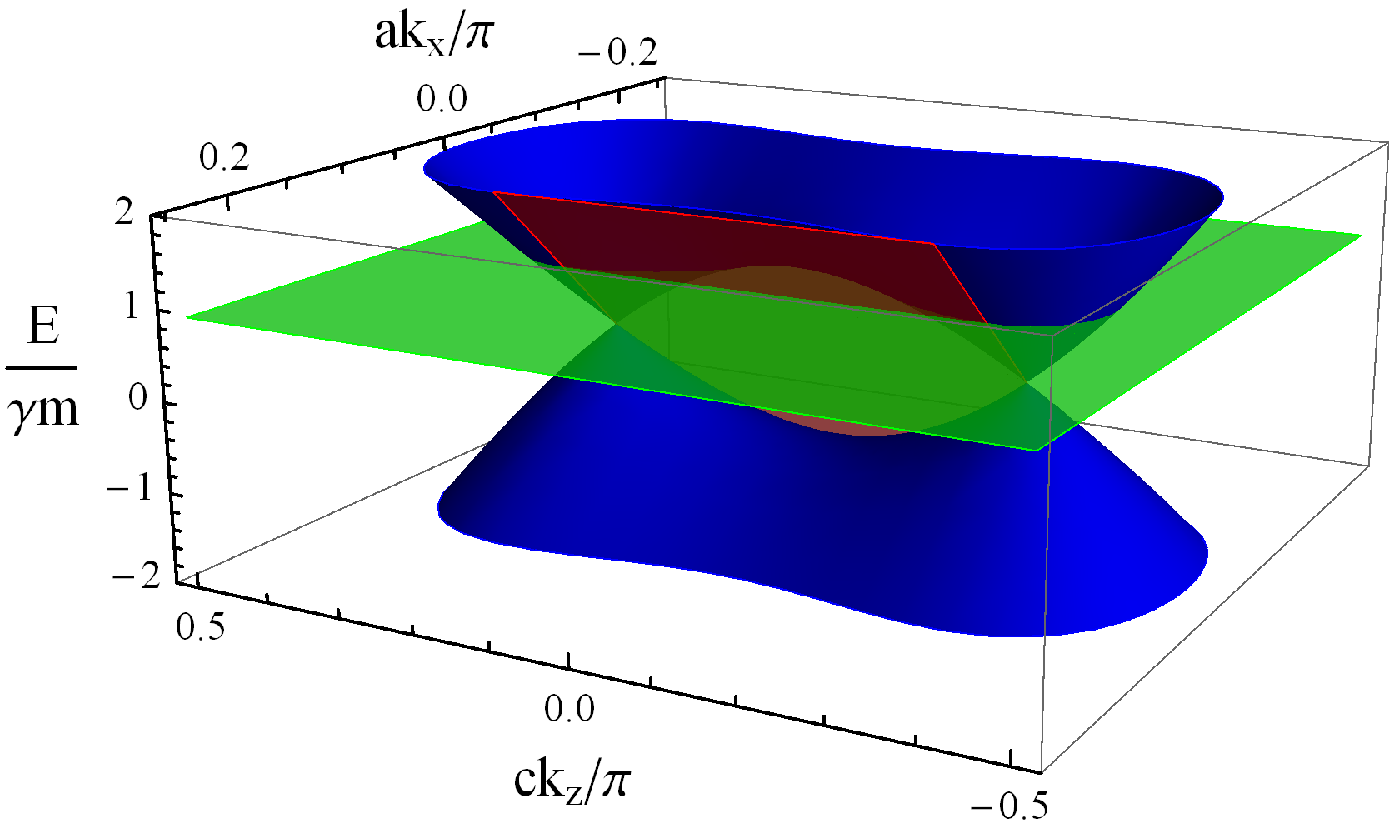}
\hspace{0.1\textwidth}
\includegraphics[width=0.4\textwidth]{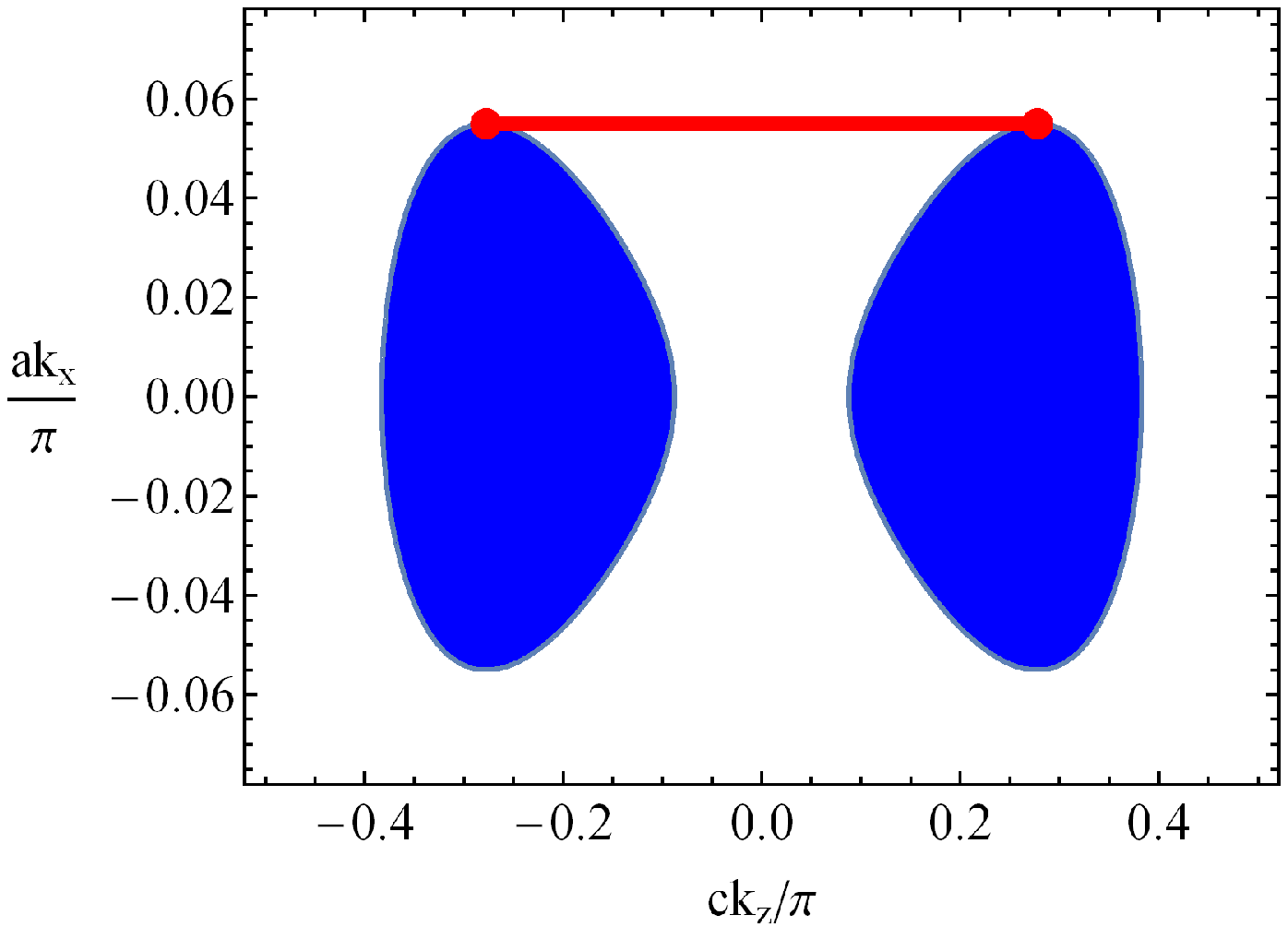}
\end{center}
\caption{The bulk (blue) and Fermi arc (red) states in the model of Weyl semimetal at $y>0$ given 
by Eq.~(\ref{Hamiltonian-effective-M1}) with $k_y=0$ (left panel). The green plane represents the 
Fermi energy $E_F/(\gamma m)=0.9$. The corresponding Fermi surface is plotted in the right panel.}
\label{fig:Fermi-surf}
\end{figure*}

The energy spectrum of model (\ref{Hamiltonian-effective-M1}) is shown in Fig.~\ref{fig:Fermi-surf}.
In order to obtain the corresponding numerical results, we chose the model parameters as in
Ref.~\cite{Wang}, i.e.,
\begin{equation}
m=0.0082\, \mbox{\AA}^{-2},\qquad \gamma=10.6424~\mbox{eV\,\AA}^2,\qquad
v_F=2.4598~\mbox{eV\,\AA},\qquad a=5.448~\mbox{\AA}, \qquad c=9.655~\mbox{\AA}.
\label{model-parameters}
\end{equation}
This choice mimics the effective low-energy description of $\mathrm{Na_3Bi}$. As we see from
Fig.~\ref{fig:Fermi-surf}, the Fermi surface consists of the bulk sheets connected by a Fermi arc.

In our study of Fermi arcs transport, we will use the simplest model of quenched disorder, in which
the interaction of electrons with impurities is described by the following Hamiltonian:
\begin{equation}
H_{\rm dis} = \int d^3 \mathbf{r}\, \psi^{\dag}(\mathbf{r})U(\mathbf{r})\psi(\mathbf{r}),
\label{H-disorder}
\end{equation}
 with the local disorder potential
\begin{equation}
U(\mathbf{r}) = \sum_j u(\mathbf{r}-\mathbf{r}_j)= \sum_j u_0 \delta(\mathbf{r}-\mathbf{r}_j),
\label{U-disorder}
\end{equation}
whose strength is determined by $u_0$.
This simple form of disorder greatly simplifies many calculations. In some cases, however,
the singular nature of the $\delta$-function may cause unphysical divergencies. When that
happens, one may replace the potential in Eq.~(\ref{U-disorder}) with a Gaussian function,
i.e.,
\begin{equation}
U(\mathbf{r}) = \sum_j u_0 \frac{e^{-\frac{(\mathbf{r}-\mathbf{r}_{j})^2}{2g^2}}}{(2\pi g^2)^{3/2}}.
\label{U-disorder-Gauss}
\end{equation}
The translation invariance, which is broken by the disorder, can be effectively restored by averaging
over the positions of impurities, i.e.,
\begin{equation}
\langle A(\mathbf{r})\rangle_{\rm dis} =\frac{1}{V} \sum_j \int d^3\mathbf{r}_j\, A(\mathbf{r}_j),
\label{disorder-average}
\end{equation}
where $V$ is the volume of the system. The impurity correlation function for the local disorder
with the potential in Eq.~(\ref{U-disorder}) is
\begin{equation}
\langle u(\mathbf{r}-\mathbf{r}_j)u(\mathbf{r}^{\prime}-\mathbf{r}_j)\rangle_{\rm dis}
=u_0^2n_{\rm imp}\delta(\mathbf{r}-\mathbf{r}^{\prime}),
\label{imp-cor-func}
\end{equation}
where $n_{\rm imp}=N_{\rm imp}/V$ is the concentration of impurities.

\section{Fermi arcs conductivity in the effective one-dimensional model}
\label{sec:LinResp-Arc-surface}

In this section, we investigate the transport properties of the surface Fermi arc states by using
the effective one-dimensional model (\ref{Hamiltonian-effective-surf-M1}). The tacit assumption
of such an approach is that the dynamics of the surface states decouples from the dynamics of
the bulk states. (As we argue later, this is not the case in Weyl semimetals.)

We will utilize the Kubo's linear response formalism. Formally, the conductivity
tensor is expressed through the retarded current-current correlation function, i.e.,
\begin{equation}
\sigma_{nm}(\omega; \mathbf{r}, \mathbf{r}^{\prime}) = -\frac{i}{\omega}
\Pi_{nm}(\omega; \mathbf{r}, \mathbf{r}^{\prime}),
\label{conductivity-definition}
\end{equation}
where the correlator itself is defined via the following ground state expectation value:
\begin{equation}
\Pi_{nm}(t-t^{\prime}; \mathbf{r}, \mathbf{r}^{\prime}) = i \theta(t-t^{\prime}) \langle [j^{n}(t, \mathbf{r}),
j^{m}(t^{\prime}, \mathbf{r}^{\prime})]\rangle .
\label{correlation-function-general}
\end{equation}
To one-loop order, the correlation function is given by the following expression:
\begin{equation}
\Pi_{xx}(i\Omega_n; \mathbf{r}, \mathbf{r}^{\prime})  =e^2v_F^2 T
\sum_{l} \langle G(i\omega_l ;\mathbf{r},\mathbf{r}^{\prime})
G(i\omega_l-i\Omega_n;\mathbf{r}^{\prime},\mathbf{r}) \rangle_{\rm dis},
\label{Pi-x-omega-M-Green}
\end{equation}
where $\omega_l =(2l+1)\pi T$ and $\Omega_n=2n\pi T$ are the fermion and boson
Matsubara frequencies, respectively. After averaging the expression in
Eq.~(\ref{Pi-x-omega-M-Green}) over the positions of impurities, we obtain
a translation invariant result $\Pi(i\Omega_n; \mathbf{r}-\mathbf{r}^{\prime})$.

\subsection{Approximation without cross-correlation diagrams}

In order to derive an approximate result for the conductivity, one might try replacing
the average $\langle G(i\omega_l ;\mathbf{r},\mathbf{r}^{\prime})
G(i\omega_l-i\Omega_n;\mathbf{r}^{\prime},\mathbf{r})\rangle_{\rm dis}$ with the
product $\langle G(i\omega_l ;\mathbf{r},\mathbf{r}^{\prime})\rangle_{\rm dis}
\langle G(i\omega_l-i\Omega_n;\mathbf{r}^{\prime},\mathbf{r})\rangle_{\rm dis}$
in the definition of the current-current correlator (\ref{Pi-x-omega-M-Green}).
In three-dimensional models, such an approximation is usually not too bad.
However, as is well known and as we will show below, it fails in the
one-dimensional model at hand. Still, such an approximation is
instructive for the later study of the full model
as a useful reference point.

By taking into account that each disorder-averaged propagator is translation
invariant, it is natural to use the momentum-space representation for the
correlation function:
\begin{equation}
\Pi_{xx}(i\Omega_n, \mathbf{q}) \simeq e^2v_F^2 T
\sum_{l} \int\frac{d^2 \mathbf{k}}{(2\pi)^2}\, \langle G(i\omega_l, \mathbf{k}) \rangle_{\rm dis}
\langle G(i\omega_l-i\Omega_n, \mathbf{k}-\mathbf{q}) \rangle_{\rm dis},
\label{Pi-q-omega-M-Green}
\end{equation}
where $\langle G(i\omega_l;\mathbf{k})\rangle_{\rm dis}$ is the Fourier transform of the coordinate-space
propagator averaged over the positions of impurities. Furthermore, it is convenient
to express the propagator in terms of its spectral function $A(\omega, \mathbf{k})$,
\begin{equation}
\langle G(i\omega_l, \mathbf{k})\rangle_{\rm dis}
=\int_{-\infty}^{\infty} d\omega \frac{A(\omega, \mathbf{k})}{i\omega_l+\mu-\omega},
\label{Green-spectral-def}
\end{equation}
where $\mu$ is the chemical potential and, by definition,
\begin{equation}
A(\omega, \mathbf{k})=\frac{i}{2\pi} \left[\langle G^{R}(\omega, \mathbf{k})\rangle_{\rm dis}-\langle G^{A}(\omega, \mathbf{k})\rangle_{\rm dis}\right]_{\mu=0} =
\frac{i}{2\pi}\left[\langle G(\omega+i0, \mathbf{k})\rangle_{\rm dis}-\langle G(\omega-i0, \mathbf{k})\rangle_{\rm dis}\right]_{\mu=0}.
\label{spectral-function-def}
\end{equation}
As usual, the retarded forms of the correlation
functions and propagators are obtained by replacing $i\omega_l\rightarrow\omega +i0$
and $i\Omega_n\rightarrow\Omega+i0$. 
In terms of the spectral function, the conductivity can be expressed as follows:
\begin{equation}
\sigma_{xx}(\Omega, \mathbf{q})
=-\frac{i}{\Omega} e^2v_F^2  \int\frac{d^2 \mathbf{k}}{(2\pi)^2}
\int \int d\omega d\omega^{\prime} \frac{ n(\omega-\mu)-n(\omega^{\prime}-\mu) }
{\omega-\omega^{\prime}-\Omega-i0}A(\omega, \mathbf{k})A(\omega^{\prime}, \mathbf{k}-\mathbf{q}),
\label{conductivity-calc-1}
\end{equation}
where $n(\omega)= 1/\left(e^{\omega/T}+1\right)$ is the Fermi-Dirac distribution function.
Note that, in the last expression, we calculated the sum over the Matsubara frequencies and performed the
appropriate analytical continuation to real values of frequencies, $i\Omega_n\rightarrow\Omega+i0$.

The expression for the dc conductivity of the Fermi arcs can now be obtained from
Eq.~(\ref{conductivity-calc-1}) by calculating the appropriate limit: $q\to 0$ and
$\Omega \to 0$. The correct order of limits is to take the long-range limit ($q\to 0$)
first and the static limit ($\Omega \to 0$) second. Then, one derives the following general
result for the real part of the dc conductivity:
\begin{equation}
\mathfrak{Re}\,\sigma_{xx}(0, 0) = -\frac{e^2v_F^2}{8T} \int \frac{d^2 \mathbf{k}}{2\pi}
\int  d\omega \frac{A(\omega, \mathbf{k})A(\omega, \mathbf{k})}{\cosh^2{\left(\frac{\omega-\mu}{2T}\right)}},
\label{conductivity-20}
\end{equation}
where we used the Sokhotski formula in order to extract the real part.

\subsection{Simple model for the quasiparticle decay width}
\label{Simple-model-width}

In the effective model (\ref{Hamiltonian-effective-surf-M1}), the explicit form of the free (retarded)
propagator for the Fermi arc states reads
\begin{equation}
G_{\rm arc}(\omega,\mathbf{k})=i\frac{\theta(m-k_z^2)}{\omega+\mu-v_Fk_x+i0}.
\label{Green-def-Ham}
\end{equation}
Notice the overall factor $\theta(m-k^2_z)$, which takes into account the fact that the arc states exist only for
a finite range of $k_z$: $-\sqrt{m}<k_z<\sqrt{m}$. The spectral function of the corresponding surface 
propagator in the clean limit is obtained by making use of the definition in Eq.~(\ref{spectral-function-def}),
\begin{equation}
A(\omega, \mathbf{k}) =i \delta(\omega-v_Fk_x)\,\theta(m-k_z^2).
\label{spectral-function-def-0}
\end{equation}
As is clear, this model describes surface Fermi arc quasiparticles with a vanishing decay width. A nonzero 
decay width $\Gamma$ caused by impurities can be introduced semi-rigorously by replacing the 
$\delta$-function in the last expression with a Lorentzian distribution,
\begin{equation}
A_{\Gamma}(\omega, \mathbf{k})=\frac{i}{\pi} \frac{\Gamma}{(\omega-v_Fk_x)^2+\Gamma^2}\,\theta(m-k_z^2).
\label{spectral-function-width}
\end{equation}
As we will argue  in Sec.~\ref{sec:width} below, it is indeed justified to introduce a nonzero decay width for the Fermi arc
quasiparticles. Here we will ignore the fact that the specific model details in Eq.~(\ref{spectral-function-width})
may not capture all realistic properties of surface quasiparticles.

In order to make a rough estimate for the quasiparticle width of the Fermi arc states, we can use
a perturbation theory in the impurity potential. In the model of quenched disorder with a local potential,
the impurity correlation function (\ref{imp-cor-func}) in momentum space is given by
\begin{equation}
D(\omega, \mathbf{k}) = 2\pi \, n_{\rm imp} u_0^2 \delta(\omega).
\label{disorder-propagator}
\end{equation}
Then, to one-loop order, the electron self-energy of the surface Fermi arc states equals
\begin{equation}
\Sigma_{\rm arc}(\Omega,\mathbf{q}) \equiv
\raisebox{-0.1\height}{\includegraphics[width=0.15\textwidth]{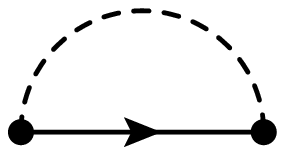}}
=- i\int \frac{d\omega d^2\mathbf{k}}{(2\pi)^{3}} G_{\rm arc}(\omega, \mathbf{k})
D(\Omega-\omega, \mathbf{q}-\mathbf{k})=- in_{\rm imp}u_0^2\int \frac{d^2\mathbf{k}}{(2\pi)^{2}} G_{\rm arc}(\Omega, \mathbf{k}).
\label{disorder-SE}
\end{equation}
By making use of Eq.~(\ref{Green-def-Ham}), we find
\begin{equation}
\Sigma_{\rm arc}(\omega,0)=-i n_{\rm imp}u_0^2 \frac{\sqrt{m}}{2\pi v_F}.
\label{disorder-SE-FA}
\end{equation}
Thus, in this approximation, the quasiparticle width is given by
\begin{equation}
\Gamma=n_{\rm imp}u_0^2 \frac{\sqrt{m}}{2\pi v_F}.
\label{Gamma-SE-FA}
\end{equation}
Finally, by making use of the general result for the dc conductivity in Eq.~(\ref{conductivity-20}),
we derive
\begin{equation}
\mathfrak{Re}\,\sigma_{xx}(0, 0) =
\frac{e^2v_F^2}{8T\pi^2} \frac{\sqrt{m}}{\pi} \int dk_x
\int  d\omega \frac{1}{\cosh^2{\left(\frac{\omega-\mu}{2T}\right)}} \frac{\Gamma^2}{\left[(\omega-v_Fk_x)^2+\Gamma^2\right]^2}
=\frac{e^2v_F \sqrt{m}}{4\Gamma\pi^2} .
\label{conductivity-2}
\end{equation}
Formally, this result for the conductivity shows that the transport is Ohmic (dissipative)
whenever the surface arc quasiparticles have a nonzero width. The nondissipative regime is
possible only when the quasiparticle width vanishes, i.e., $\Gamma\to 0$. Therefore, the issue
of the quasiparticle width for the surface states appears to be an important aspect that can shed
light on the nature of the arc's transport. We will try to resolve this issue later. Before doing that,
however, we should also address the validity of the approximation for the current-current
correlator with cross-correlation diagrams neglected. As is well known and as we show below, such an approximation
is not reliable. It will be essential, therefore, to account for some cross-correlations. This will be
done via the inclusion of the appropriate vertex corrections in Sec.~\ref{sec:vertex}.

\subsection{Nondissipative transport from nonperturbative treatment in one dimension}
\label{sec:LinResp-Arc-surface-2}

It is easy to show that the naive calculation of the conductivity in the previous subsection
is unreliable for one-dimensional chiral fermions, described by the effective low-energy
model (\ref{Hamiltonian-effective-surf-M1}). The corresponding argument is well known
\cite{Girvin}.

In the presence of quenched disorder in one-dimension, the eigenstate wave-functions
for chiral fermions can be calculated exactly (note the absence of backscattering)
\begin{equation}
\psi_E(x)= e^{\frac{i}{v_F}\left[Ex- \sum_j u_0 \int_{x_0}^{x}dz\delta(z-x_j)\right]},
\label{disorder-arc-function}
\end{equation}
leading to the following exact Green's function for the corresponding Fermi arcs states:
\begin{equation}
G(\omega; x, x^{\prime})=i\int\frac{dE}{2\pi} \frac{1}{\omega+\mu-E} e^{\frac{i}{v_F}
\left[E (x-x^{\prime})- \sum_j u_0
\left(\theta(x-x_j)-\theta(x^{\prime}-x_j)\right)\right]}.
\label{Green-def-dis-calc}
\end{equation}
As we see from this nonperturbative result, the disorder potential appears only as a single phase
$\sum_j u_0 \left[\theta(x-x_j)-\theta(x^{\prime}-x_j)\right]$ in the exponent. Moreover, the
phase is such that it changes the overall sign under the interchange of $x$ and $x^{\prime}$. Therefore,
after substituting propagator (\ref{Green-def-dis-calc}) into the expression for the retarded
current-current correlation function (\ref{Pi-x-omega-M-Green}), we find that all effects of
disorder drop out from the final result. In other words, in the effective one-dimensional surface
model defined by Eq.~(\ref{Hamiltonian-effective-surf-M1}),
disorder does not affect the Fermi arc transport, i.e., the transport is nondissipative.
One should note, however, that this conclusion does not necessarily imply the vanishing
quasiparticle width for the surface states. The corresponding width $\Gamma$ is a characteristic
of the propagator (\ref{Green-def-dis-calc}) averaged over the disorder. Irrespective of the actual
value of such a width, it does not affect the result for the correlation function (\ref{Pi-x-omega-M-Green}).
This is in contrast to the naive arguments of Sec.~\ref{Simple-model-width}, which relied on the
approximation where the cross-correlation between the quasiparticle propagators was
neglected. This simple observation is instructive and should be kept in mind in the analysis
of transport properties of the Fermi arc states in a more realistic model of a Weyl (semi-)metal
in the next section.

\section{Perturbative signs of dissipation in the full model}
\label{sec:Perturbative}

As we showed in the previous section, the absence of dissipation in the Fermi arc transport in the
one-dimensional effective model of chiral fermions is connected with its very unique property that
all effects of quenched disorder in the exact Green's function $G(\omega; x, x^{\prime})$ are captured
in a single phase factor, which is odd under the interchange of $x$ and $x^{\prime}$. Because of
its special property, the phase factor drops out of the expression for the current-current correlator,
i.e., disorder has no effect on the transport.

Below, we will argue that the full effective model of Weyl semimetals (\ref{Hamiltonian-effective-M1}),
which necessarily contains both the surface Fermi arc states and the bulk states, does not
share this property. The formal reasons for this are (i) the loss of one-dimensional kinematic
constraint in scattering of chiral fermions and (ii) the nondecoupling of the low-energy dynamics
of the two types of states (i.e., surface and bulk). From a physical viewpoint, this immediately leads
to a dissipation because the surface Fermi arc states can easily scatter into the bulk. With the
loss of exact integrability in the full model, of course, we cannot present an exact solution
supporting this claim. Instead, we will use physical arguments that strongly support
nondecoupling and dissipation.

Let us consider the scattering of the Fermi arc states using the Born approximation.
A convenient starting point for such an analysis is the Lippmann--Schwinger equation
for the surface states,
\begin{equation}
\psi_s(\mathbf{r})=\psi^{(0)}_s(\mathbf{r})-i\int d^3\mathbf{r}^{\prime}
S(\mathbf{r}, \mathbf{r}^{\prime})U(\mathbf{r}^{\prime})
\psi_s(\mathbf{r}^{\prime} ),
\label{LP-equation}
\end{equation}
where $\psi^{(0)}_s(\mathbf{r})\equiv\psi_{y>0}$ is the incident surface wave as in
Eq.~(\ref{psi-surface-states}). Strictly speaking, here we should use a regularized
Gaussian form of the disorder potential $U(\mathbf{r}^{\prime})$, given in
Eq.~(\ref{U-disorder-Gauss}). Indeed, if a point-like interaction is used, the
second term in the Lippmann--Schwinger equation would give a contribution
to the wave-function $\sum_{j} u_0 S(\mathbf{r}, \mathbf{r}_{j}) \psi_s(\mathbf{r}_{j})$,
which diverges at the location of impurities.

In general, solving the Lippmann--Schwinger equation is a quite difficult (nonperturbative) problem.
Instead, here we will consider a perturbative solution in the Born approximation, in which
the surface state wave-function $\psi_s(\mathbf{r}^{\prime})$ on the right hand side of Eq.~(\ref{LP-equation}) is replaced with
$\psi^{(0)}_s(\mathbf{r}^{\prime})$.

It may be instructive to recall that the Born approximation can be readily
derived from the spectral problem, $(H_0+U)\psi = E\psi$. Assuming a perturbative expansion of
the solution in the form $\psi \simeq \psi^{(0)} +\psi^{(1)}$, where $\psi^{(0)}$ are the eigenstates
of the free Hamiltonian $H_0$, we easily derive the formal result for the first-order correction:
$\psi^{(1)} = (E-H_0)^{-1} U \psi^{(0)}$. Noting that $S \equiv i(E-H_0)^{-1}$ is the operator form
of the free propagator and taking into account its coordinate-space representation, we see that
$\psi^{(1)}$ coincides with the second term in Eq.~(\ref{LP-equation}) in the Born approximation.

\subsection{Free propagator in the full model}
\label{sec:freePropagator}

In the clean limit (i.e., no disorder), it is straightforward to derive an explicit expression for
the retarded quasiparticle Green's function (free propagator) in the full effective model
(\ref{Hamiltonian-effective-M1}) by making use of the general definition:
$S^{R}(\omega,\mathbf{r}, \mathbf{r}^{\prime})\equiv i\sum_E \psi_E(\mathbf{r})
\psi_E^{\dagger}(\mathbf{r}^{\prime})/(\omega-E+i0)$,
where the sum runs over a complete set of the energy eigenstates.
By taking into account that the Hilbert space of the corresponding eigenvalue
problem contains both bulk and surface states, the resulting Green's function
naturally splits into the surface and bulk contributions:
\begin{equation}
S^{R}(\omega,\mathbf{r}, \mathbf{r}^{\prime}) =\int \frac{d^2\mathbf{k}_{\parallel}\,
e^{i\mathbf{k}_{\parallel}(\mathbf{r}_{\parallel}-\mathbf{r}_{\parallel}^{\prime})}}{(2\pi)^2}\,
S^{R}_{s}(\omega, \mathbf{k}_{\parallel}; y, y^{\prime}) + \int \frac{d^3\mathbf{k}\,\,
e^{i\mathbf{k}(\mathbf{r}-\mathbf{r}^{\prime})}}{(2\pi)^3}\,S^{R}_{b}(\omega, \mathbf{k}).
\label{G-def}
\end{equation}
In the model at hand, the explicit forms of the two types of the Green's functions are
\begin{eqnarray}
\label{S-surf}
S^{R}_{s}(\omega, \mathbf{k}_{\parallel}; y, y^{\prime}) &=&
ip(k_z)\frac{\left(1+\sigma_x\right) e^{-(y+y^{\prime})p(k_z)}}{\omega-v_Fk_x+i0} , \\
\label{S-bulk}
S^{R}_{b}(\omega, \mathbf{k})&=& i\frac{\omega +\gamma(k_z^2-m)\sigma_z
+v_F(k_x\sigma_x+k_y\sigma_y)}{\omega^2-E(k)^2+i0\sign{(\omega)}},
\end{eqnarray}
where $E(k)=\pm\sqrt{\gamma^2(k_z^2-m)^2+v_F^2(k_x^2+k_y^2)}$. Here and in what follows 
we assume that $-\sqrt{m}<k_z<\sqrt{m}$ in all surface functions. (Strictly 
speaking, the Green's function of bulk states is not translation invariant in the presence of the 
surface at $y=0$. However, this is relevant only near the surface and we checked that taking 
into account the translation non-invariance of the bulk Green's function does not affect our
qualitative conclusions. Therefore,
for the sake of simplicity, we use the translation invariant Green's function of the bulk states.)
Note that the coordinate-space representation of the surface Green's function of the Fermi
arc states is given by
\begin{equation}
S^{R}_{s}(\omega;\mathbf{r}, \mathbf{r}^{\prime})
= \frac{\theta\left(x-x^{\prime}\right)e^{\frac{i\omega(x-x^{\prime})}{v_F}}}{2\pi v_F}
\left(1+\sigma_x\right) F_{\mathbf{r}, \mathbf{r}^{\prime}},
\label{S-surf-coord}
\end{equation}
where
\begin{eqnarray}
F_{\mathbf{r}, \mathbf{r}^{\prime}}&=& \frac{1}{8\sqrt{v_F} \gamma^{3/2} (y+y^{\prime})^{5/2}} \Bigg\{-4\sqrt{\gamma v_F (y+y^{\prime})} \left[v_F(z-z^{\prime})\sin{(\sqrt{m}(z-z^{\prime}))} +2\gamma \sqrt{m}(y+y^{\prime}) \cos{(\sqrt{m}(z-z^{\prime}))}\right] \nonumber \\
&+&i\sqrt{\pi}e^{\frac{v_F(z-z^{\prime})^2}{4\gamma (y+y^{\prime})} -\frac{\gamma m (y+y^{\prime})}{v_F}} \left[ \mathrm{Erf}\left(\frac{v_F(z-z^{\prime}) -2i\sqrt{m}\gamma (y+y^{\prime})}{2\sqrt{v_F\gamma (y+y^{\prime})}}\right) -\mathrm{Erf}\left(\frac{v_F(z-z^{\prime}) +2i\sqrt{m}\gamma (y+y^{\prime})}{2\sqrt{v_F\gamma (y+y^{\prime})}}\right)\right] \nonumber\\ &\times&\left[v_F^2(z-z^{\prime})^2+2v_F\gamma(y+y^{\prime})+4m\gamma^2(y+y^{\prime})^2\right]\Bigg\},\label{F-surf}
\end{eqnarray}
and $\mathrm{Erf}(z)= \frac{2}{\sqrt{\pi}}\int_0^{z}e^{-t^2}dt$ is the error function \cite{Stegun}. By making use of the
well-known asymptotic behavior of the error function (see formula 7.1.23 in Ref.~\cite{Stegun}), we
derive the following results:
\begin{eqnarray}
\label{F-surf-asym-z}
F_{\mathbf{r}, \mathbf{r}^{\prime}} &\simeq & -4 \sqrt{m} \gamma
\frac{\cos{\left(\sqrt{m}(z-z^{\prime})\right)}}{v_F(z-z^{\prime})^2} +\mathcal{O}\left(\frac{1}{(z-z^{\prime})^3}\right) ,
\quad \mbox{for} \quad z \to \infty ,\\
\label{F-surf-asym-y}
F_{\mathbf{r}, \mathbf{r}^{\prime}} &\simeq& \frac{v_F \cos{\left(\sqrt{m}(z-z^{\prime})\right)}}{\sqrt{m}\gamma (y+y^{\prime})^2} +\mathcal{O}\left(\frac{1}{(y+y^{\prime})^3}\right), \quad \mbox{for} \quad y \to \infty .
\end{eqnarray}
Interestingly, we find that $F_{\mathbf{r}, \mathbf{r}^{\prime}}$ falls off as a power-law function of $y$, rather than
an exponential function, when $y\to \infty$. At first sight, this seems surprising since (almost)
all surface states, see Eq.~(\ref{psi-surface-states}), have an exponential dependence on $y$.
The puzzle is resolved by noting that the exponents vanish near the ends of the Fermi arcs,
i.e., when $k_z\to \pm \sqrt{m}$. Thus, when integrating over all momenta in the surface
Green's function, the contributions from the regions with a (nearly) vanishing exponent
dominate the coordinate-space asymptotes. In fact, this is also an indication
that the dynamics of the surface arc states in the full model does not decouple from the
dynamics of the bulk states.

\subsection{Scattering of Fermi arc states in the Born approximation}
\label{sec:Born}

By making use of the general result in the Born approximation and the explicit form of the free
propagator (\ref{G-def}), to the first order in the disorder potential, the scattered wave
is given by
\begin{eqnarray}
\label{psi-1}
\psi^{(1)}(\mathbf{r}) &=& \psi^{(1)}_s(\mathbf{r})+\psi^{(1)}_b(\mathbf{r}) , \\
\label{psi-1-s}
\psi^{(1)}_s(\mathbf{r}) &\simeq & -i\int d^3\mathbf{r}^{\prime} S_{s}^{R}(\mathbf{r}, \mathbf{r}^{\prime})
U(\mathbf{r}^{\prime})\psi^{(0)}_s(\mathbf{r}^{\prime} )
\simeq -i  u_{0} \sum_{j} S_{s}^{R}(\mathbf{r}, \mathbf{r}_{j})  \psi^{(0)}_s(\mathbf{r}_{j} ), \\
\label{psi-1-b}
\psi^{(1)}_b(\mathbf{r}) &\simeq & -i\int d^3\mathbf{r}^{\prime} S_{b}^{R}(\mathbf{r}, \mathbf{r}^{\prime})
U(\mathbf{r}^{\prime})\psi^{(0)}_s(\mathbf{r}^{\prime} )
\simeq -i  u_{0} \sum_{j} S_{b}^{R}(\mathbf{r}, \mathbf{r}_{j})  \psi^{(0)}_s(\mathbf{r}_{j} ).
\end{eqnarray}
Note that there are separate surface and bulk waves. This is in agreement with the structure
of the quasiparticle Green's function in the full model, which contains both the surface and bulk
contributions, see Eqs.~(\ref{G-def}) through (\ref{S-bulk}). As mentioned earlier, a regularized
version of the impurity potential (\ref{U-disorder-Gauss}) should be used in the above expression
for the scattered bulk wave. Otherwise, the result will be singular at $\mathbf{r}\to \mathbf{r}_{j}$.
For the purposes of extracting the outgoing waves at large distances $r\to\infty$, however, it suffices
to use the local form of the potential given by Eq.~(\ref{U-disorder}).

Then, by making use of the propagator in Eq.~(\ref{S-surf}) and the $z\to \infty$ asymptote from
Eq.~(\ref{F-surf-asym-z}), we derive the following result for the surface part of the scattered wave:
\begin{equation}
\psi_s^{(1)}(\mathbf{r}) \simeq
4 i\sqrt{m} \gamma u_0 \sum_{j} \frac{\theta\left(x-x_{j}\right) \cos\left[\sqrt{m}(z-z_{j})\right] }{\pi v_F^2 (z-z_{j})^2}
\sqrt{p(k_z)}  e^{ik_xx+ik_zz_{j} -p(k_z)y_{j}}\left(
                                                              \begin{array}{c}
                                                                1 \\
                                                                1 \\
                                                              \end{array}
                                                            \right).
\label{born-surface-local}
\end{equation}
While the $x$ component of momentum $k_x$ remains unchanged in the scattered wave, the
$z$ component of momentum drastically changes from the original value $k_z$, constrained
only by its range: $-\sqrt{m} <k_z<\sqrt{m}$. Indeed, the outgoing wave is proportional to 
$\cos\left[\sqrt{m}(z-z_{j})\right]$, i.e., it is a superposition of two waves with
limiting values of the momenta, $k_z=\pm \sqrt{m}$. Since the distinction between the two types of
states fades away in the corresponding region of the phase space, one might argue that the Fermi
arc states are pushed towards the bulk states.

\begin{figure*}[!ht]
\begin{center}
\includegraphics[width=0.32\textwidth]{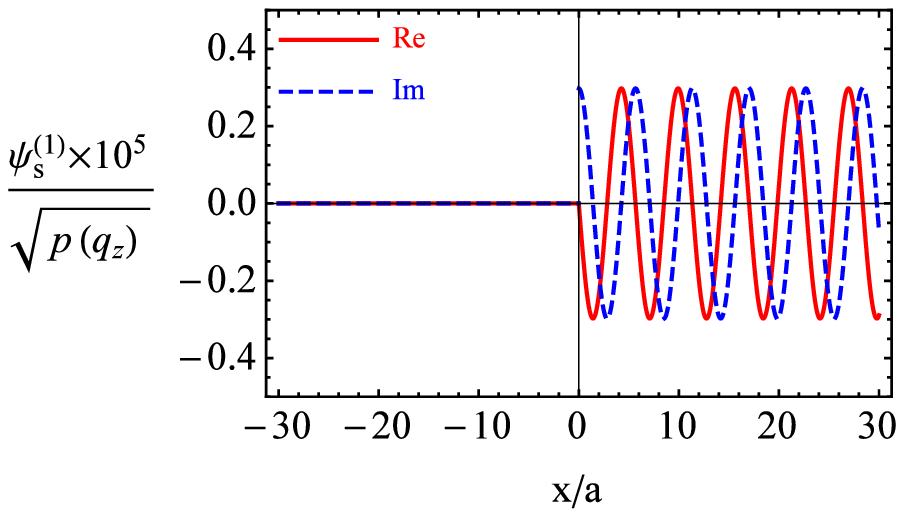}
\hfill
\includegraphics[width=0.32\textwidth]{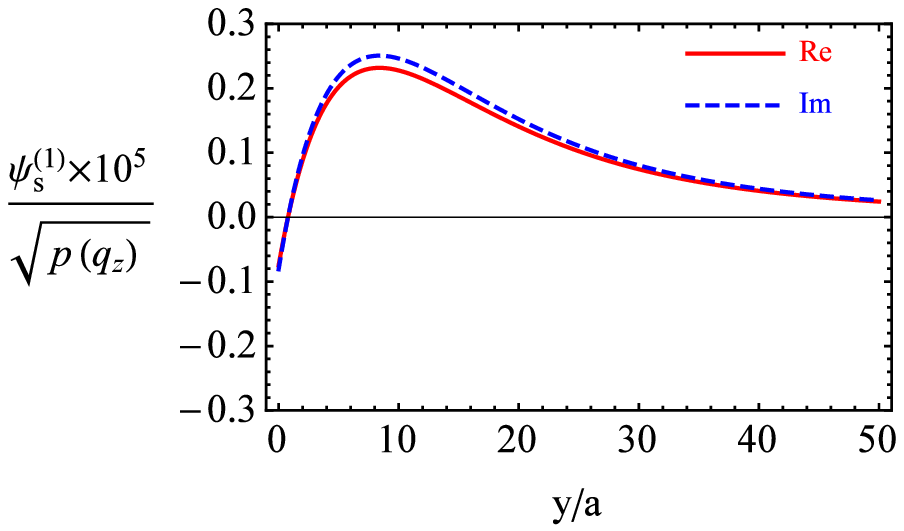}
\hfill
\includegraphics[width=0.32\textwidth]{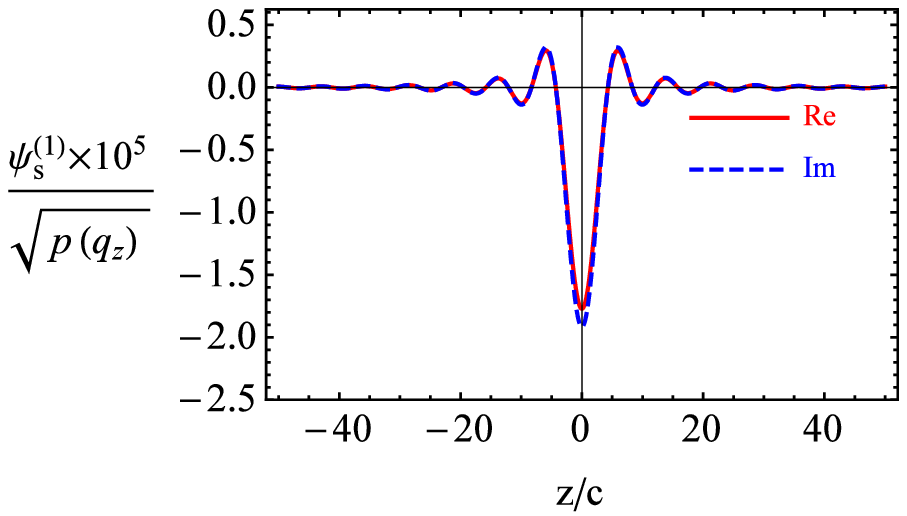}
\end{center}
\caption{(Color online) The upper component of the surface part of scattered wave $\psi_s^{(1)}$,
defined by Eq.~(\ref{born-surface-local}), as a function of $x/a$ (left panel), $y/a$ (middle panel),
and $z/c$ (right panel). The numerical values of the model parameters are
$u_0=0.1~\mbox{eV\, \AA}^{3}$, $\omega=0.5~\mbox{eV}$, $\mathbf{q}=(0,0,0)$,
$\mathbf{r}_j=\left(\omega/v_F ,0,0\right)$, and $a$ and $c$ are defined in Eq.~(\ref{model-parameters}).
The wave function is shown for
$\mathbf{r}=\left(x, 5a, 5c\right)$ in the left panel,
$\mathbf{r}=\left(5a, y, 5c\right)$ in the middle panel, and
$\mathbf{r}=\left(5a, 5a, z\right)$ in the right panel.}
\label{fig:Born-surface-local-all}
\end{figure*}

In the case of a single impurity, the representative numerical results for the scattered surface portion of the
wave-function, $\psi_s^{(1)}$, are shown in Fig.~\ref{fig:Born-surface-local-all}. As one can see from the
$x$-dependence in the left panel, there is no backscattering (i.e., into the region $x<0$) of the surface wave.
In the region $x>0$, the $\pi/2$ phase shift between the real and imaginary parts of $\psi_s^{(1)}$ suggests
a characteristic behavior, $e^{i k_x x}$, of a plane wave propagating forward. The other two panels
in Fig.~\ref{fig:Born-surface-local-all} show the profiles of the real and imaginary parts of the scattered wave in the $y$
direction (into the bulk) and the $z$ direction (along the surface). These latter two are of no particular significance
by themselves. However, they will serve as a useful benchmark later in our comparison with the waves
scattered into the bulk.

In order to calculate the bulk part of the scattered wave (\ref{psi-1-b}), we use the bulk propagator (\ref{S-bulk})
in momentum space and derive
\begin{equation}
\psi_b^{(1)}(\mathbf{r}) =
u_{0} \sum_{j}  \int \frac{d^3 \mathbf{k}}{(2\pi)^3}  \frac{ e^{i\mathbf{k}(\mathbf{r}-\mathbf{r}_{j}) } }
{\omega^2-E(k)^2+i0\sign{(\omega)}}
\left[\omega+\gamma(k_{z}^2-m)\sigma_z+v_F(k_{x}\sigma_x+k_{y}\sigma_y)\right] \psi^{(0)}_s( \mathbf{r}_{j}) .
\label{born-bulk-Gauss-000}
\end{equation}
In this expression, two out of three momentum integrations can be performed analytically. Then, the result reads
\begin{eqnarray}
\psi_b^{(1)}(\mathbf{r}) &=& -u_{0} \sum_{j}  \int \frac{dk_z}{4\pi^2 v_F^2}   e^{ik_z (z-z_{j}) }   \Bigg\{
\left[\omega+\gamma(k_{z}^2-m)\sigma_z\right] \left[\theta(-\Omega^2) K_{0}\left(\frac{\rho|\Omega|}{v_F} \right)
+i\frac{\pi}{2}\theta(\Omega^2) H_{0}^{(1)}\left(\sign{(\omega)}\frac{\rho|\Omega|}{v_F} \right)\right]
\nonumber \\
&+&i\frac{|\Omega|}{\rho}\left[(x-x_{j})\sigma_x+(y-y_{j})\sigma_y \right]
\left[ \theta(-\Omega^2) K_{1}\left(\frac{\rho|\Omega|}{v_F} \right)
+i\sign{(\omega)}\frac{\pi}{2}\theta(\Omega^2) H_{1}^{(1)}\left(\sign{(\omega)}\frac{\rho|\Omega|}{v_F} \right)
 \right]
\Bigg\}\psi^{(0)}_s( \mathbf{r}_{j}), \nonumber\\
\label{born-bulk-local}
\end{eqnarray}
where, by definition, $\Omega^2 \equiv  \omega^2-\gamma^2(k_z^2-m)^2$ and $\rho^2 \equiv  (x-x_j)^2+(y-y_j)^2$.

\begin{figure*}[!ht]
\begin{center}
\includegraphics[width=0.32\textwidth]{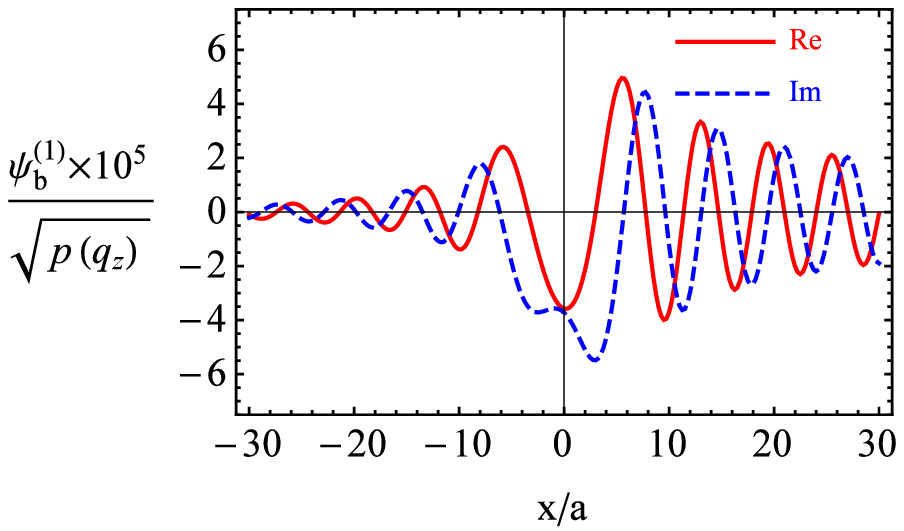}
\hfill
\includegraphics[width=0.32\textwidth]{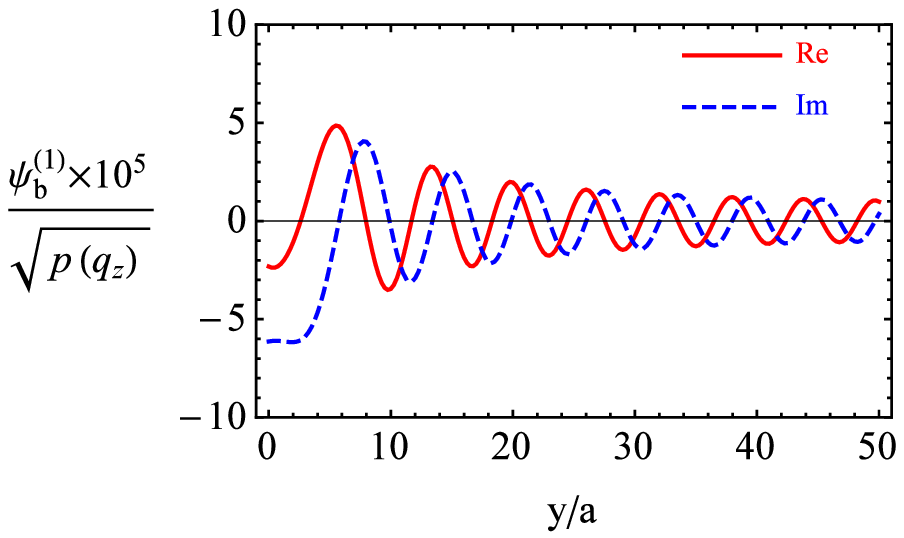}
\hfill
\includegraphics[width=0.32\textwidth]{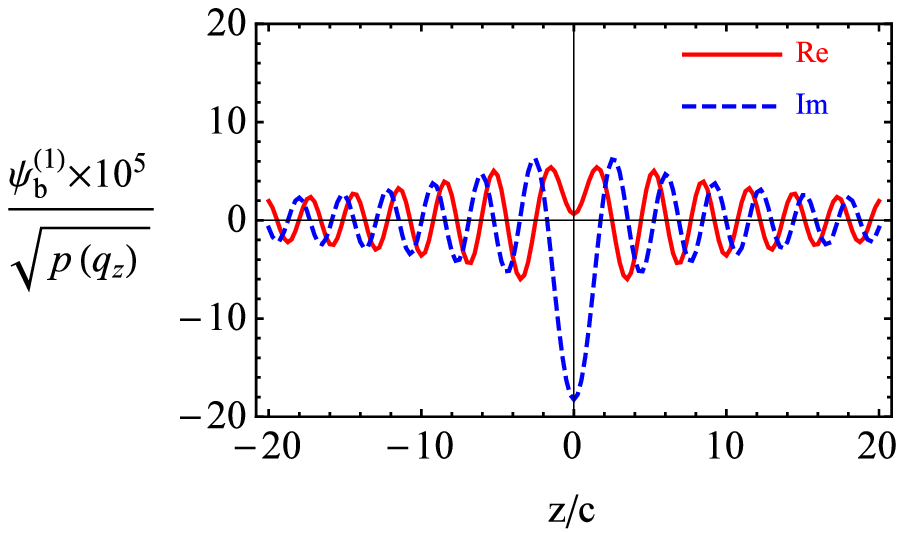}
\end{center}
\caption{(Color online) The upper component of the bulk part of scattered wave $\psi_b^{(1)}$, given by Eq.~(\ref{born-bulk-local}), as a function of $x/a$ (left panel), $y/a$ (middle panel), and $z/c$ (right panel).
The numerical values of the model parameters are
$u_0=0.1~\mbox{eV\, \AA}^{3}$, $\omega=0.5~\mbox{eV}$, $\mathbf{q}=(\omega/v_F ,0,0)$,
$\mathbf{r}_j=\left(0,0,0\right)$. The wave function is shown for
$\mathbf{r}=\left(x, 5a, 5c\right)$ in the left panel,
$\mathbf{r}=\left(5a, y, 5c\right)$ in the middle panel, and
$\mathbf{r}=\left(5a, 5a, z\right)$ in the right panel.}
\label{fig:Born-bulk-local-Up}
\end{figure*}

\begin{figure*}[!ht]
\begin{center}
\includegraphics[width=0.32\textwidth]{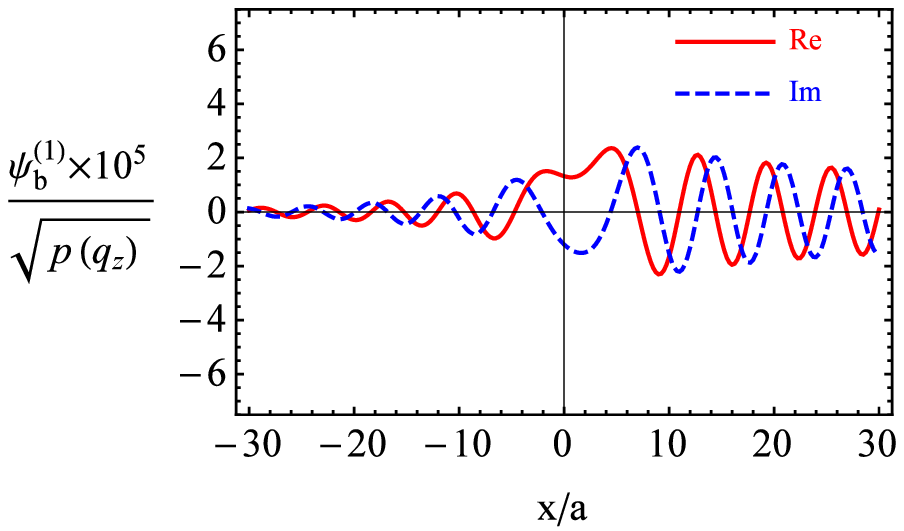}
\hfill
\includegraphics[width=0.32\textwidth]{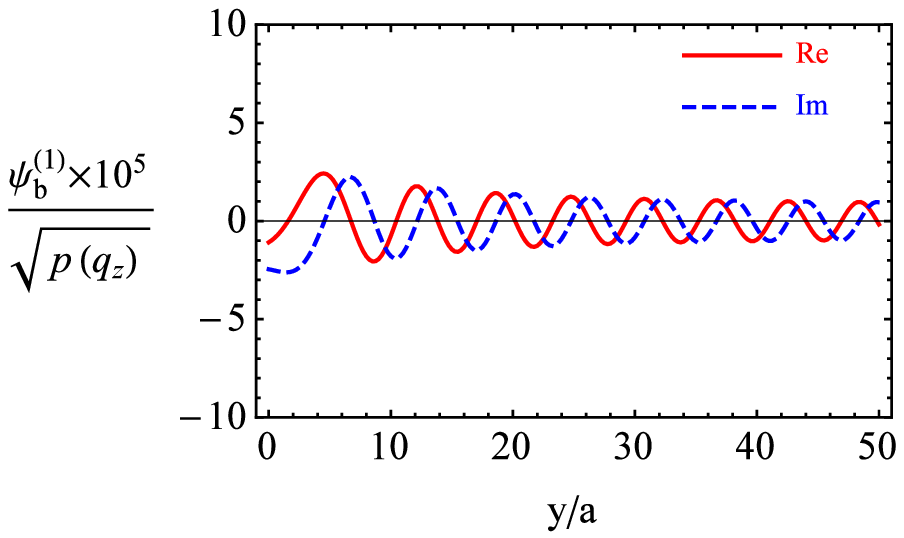}
\hfill
\includegraphics[width=0.32\textwidth]{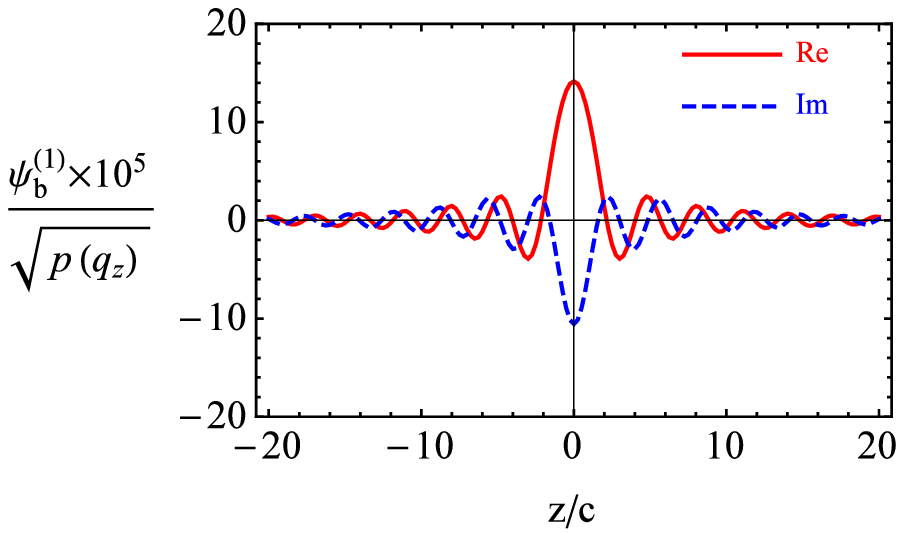}
\end{center}
\caption{(Color online) The lower component of the bulk part of scattered wave $\psi_b^{(1)}$, given by Eq.~(\ref{born-bulk-local}), as a function of $x/a$ (left panel), $y/a$ (middle panel), and $z/c$ (right panel).
The numerical values of the model parameters are
$u_0=0.1~\mbox{eV\, \AA}^{3}$, $\omega=0.5~\mbox{eV}$, $\mathbf{q}=(\omega/v_F ,0,0)$,
$\mathbf{r}_j=\left(0,0,0\right)$. The wave function is shown for
$\mathbf{r}=\left(x, 5a, 5c\right)$ in the left panel,
$\mathbf{r}=\left(5a, y, 5c\right)$ in the middle panel, and
$\mathbf{r}=\left(5a, 5a, z\right)$ in the right panel.}
\label{fig:Born-bulk-local-Down}
\end{figure*}

In the case of a single impurity, the numerical results for the scattered bulk portion of the wave-function,
$\psi_b^{(1)}$, are shown in Figs.~\ref{fig:Born-bulk-local-Up} and \ref{fig:Born-bulk-local-Down}. From
the $x$-dependence of the scattered wave, presented in the left panels of Figs.~\ref{fig:Born-bulk-local-Up}
and \ref{fig:Born-bulk-local-Down}, we see clearly that backscattering is present. This is in contrast
to the surface scattered wave, studied earlier. There are also qualitative differences in the $z$ and
$y$ dependence of the scattered bulk wave, presented in the middle and right panels of
Figs.~\ref{fig:Born-bulk-local-Up} and \ref{fig:Born-bulk-local-Down}, respectively. Unlike the corresponding
surface counterparts, studied earlier, these results reveal outgoing waves propagating into the bulk.
This is suggested by the characteristic phase shift of $\pi/2$ between the real and imaginary parts of
all components of the wave-function.


\begin{figure*}[!ht]
\begin{center}
\includegraphics[width=0.45\textwidth]{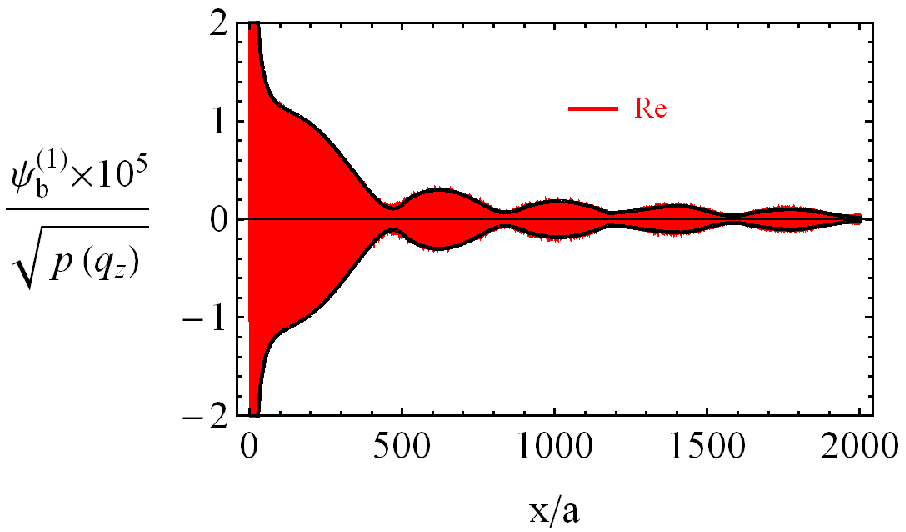}
\hfill
\includegraphics[width=0.45\textwidth]{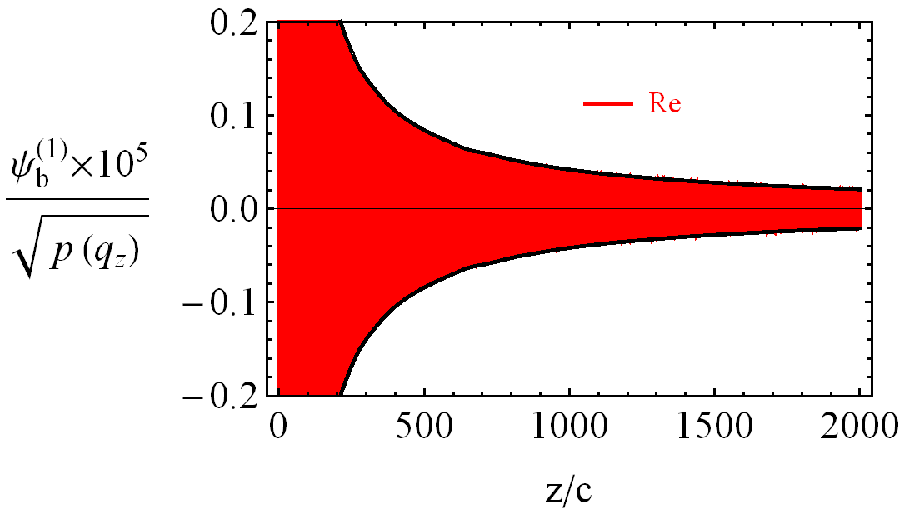}
\end{center}
\caption{(Color online) The real part of the upper component of the bulk part of scattered wave
$\psi_b^{(1)}$, given by Eq.~(\ref{born-bulk-local}), as a function of $x/a$ (left panel) and as a
function of $z/c$ (right panel) for large values of arguments. The numerical values of the model parameters 
are $u_0=0.1~\mbox{eV\, \AA}^{3}$, $\omega=0.5~\mbox{eV}$, $\mathbf{q}=(\omega/v_F ,0,0)$,
$\mathbf{r}_j=\left(0,0,0\right)$. For plotting the results, we used $\mathbf{r}=\left(x, 5a, 5c\right)$
in the left panel and $\mathbf{r}=\left(5a, 5a, z\right)$ in the right panel.}
\label{fig:Born-bulk-local-x-asym}
\end{figure*}

It is interesting to study the asymptotes of the bulk scattered wave $\psi_b^{(1)}(\mathbf{r})$ at large
values of the coordinates along the surface of the semimetal, i.e., $x\gg a$ and $z\gg c$. Both
asymptotes reveal a long-range (propagating) tail with a weak modulation of the amplitudes. This
is demonstrated in Fig.~\ref{fig:Born-bulk-local-x-asym}, in which we show the real part of the upper
component of $\psi_b^{(1)}$ as a function of $x/a$ (left panel) and as a function of $z/c$ (right panel) 
for large values of the arguments. Note, that the red shaded area in the plot is an outcome of many
oscillations with a small period. From the numerical result in Fig.~\ref{fig:Born-bulk-local-x-asym},
we find that the amplitude of the large scale modulations decreases as $1/x$ at large $x/a$ and
as $1/z^2$ at large $z/c$. As one can check, the same qualitative behavior is valid for the real part
of the lower component of $\psi_b^{(1)}$, as well as for the imaginary parts of both components.

To summarize our results, we find that, after scattering on an impurity, a surface wave produces
not only (forward-scattered) surface states, but also waves propagating into the bulk. This implies
that a naive low-energy consideration of the surface Fermi arc states is invalid. Indeed, such
approximately one-dimensional states are not decoupled from the three-dimensional bulk states.
In the context of transport properties, this means that a proper analysis of the dynamics with a
mixing of the two types of states is required. This problem is addressed in the next section.

\section{Fermi arc width and conductivity in the full model}
\label{sec:Arc-Width-Conductivity-Full}

In this section, we study the transport properties of the Fermi arcs in the full model
(\ref{Hamiltonian-effective-M1}) that contains not only the surface states, but
also the gapless bulk states. As expected, the exact integrability of the full model is
lost and we have to rely on some approximations. By making use of the simplified
analysis in the previous section, we will work out an approximate scheme that captures
the essential details of the low-energy dynamics.

The nonperturbative arguments of Sec.~\ref{sec:LinResp-Arc-surface-2} may suggest that the transport
due to the Fermi arcs is nondissipative. However, it is quite reasonable to suspect that the
description of the Fermi arc states using the effective Hamiltonian (\ref{Hamiltonian-effective-surf-M1})
is inadequate. The point is that the corresponding approach completely ignores the possibility of
the surface states scattering into the bulk. From a physics viewpoint, however, such scattering is
unavoidable at least in certain regions of the phase space. In particular, this is the case for the
surface states near the ends of the Fermi arcs, where they smoothly transform into the bulk states.
As we argue below, the resulting non-decoupling of the surface and bulk sectors in the full theory is
critical. It is responsible for dissipative transport properties.

The general definition of the conductivity in terms of the current-current correlation function
(\ref{conductivity-definition}) remains valid also in the full model (\ref{Hamiltonian-effective-M1}).
However, one has to modify the definition of the currents to account for the correct
matrix structure of the vertices, i.e.,
\begin{eqnarray}
j_{x} &=& -ev_F\psi^{\dag}\sigma_x\psi,
\label{current-Fermi-arc-x} \\
j_{z} &=& - 2e\gamma k_z\psi^{\dag}\sigma_z\psi.
\label{current-Fermi-arc-z}
\end{eqnarray}
By taking into account the matrix structure of the vertices in these currents and the matrix structure
of the surface state propagator (\ref{S-surf}), it is straightforward to show that two out of three possible
components of the conductivity tensor are vanishing, i.e., $\sigma_{xz}=\sigma_{zz}=0$. The remaining
nontrivial component of the dc conductivity tensor reads
\begin{equation}
\sigma_{xx} (\mathbf{r}) = -\lim_{\Omega \to 0} \frac{i}{\Omega} \Pi_{xx}(\mathbf{r}).
\label{vertex-conductivity-T0}
\end{equation}
Note that the result may depend on the position vector $\mathbf{r}$ (e.g., see Ref.~\cite{Rodionov:2015}).
In the coordinate space, the corresponding correlator is given by
\begin{equation}
\Pi_{xx}(\mathbf{r}) \simeq e^2v_F^2 \Omega\int\frac{d\omega}{2\pi}
\frac{\partial n(\omega-\mu)}{\partial \omega}\int d^3\mathbf{r}^{\prime}  \mathrm{tr}\left[ \langle\sigma_x
(-i)G^A_s(\omega; \mathbf{r}, \mathbf{r}^{\prime})\sigma_x
(-i)G^R_s(\omega; \mathbf{r}^{\prime}, \mathbf{r}) \rangle_{\rm dis}\right],
\label{vertex-correlator}
\end{equation}
where we kept only the linear term in $\Omega$, which is relevant for the dc conductivity. In the last
expression, $n(\omega-\mu)$ is the Fermi-Dirac quasiparticle distribution function. 

It is understood that the full (advanced and retarded) propagators $G^{A/R}_s(\omega; \mathbf{r}, \mathbf{r}^{\prime})$
are calculated in the presence of disorder. The corresponding expressions could be formally obtained
in the form of infinite series of diagrams with different number of insertions, representing scatterings
on impurities, i.e.,
\begin{equation}
G \equiv \raisebox{-0.05\height}{\includegraphics[width=0.5\textwidth]{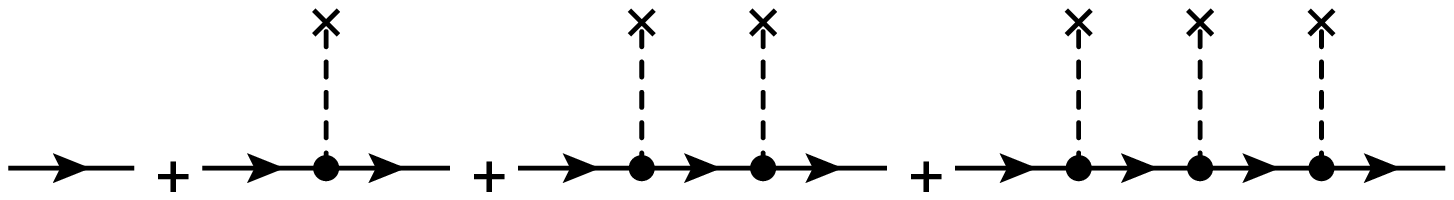}} + \cdots
\label{diagram:propagator}
\end{equation}
After substituting the propagators into the expression for the correlation function (\ref{vertex-correlator})
and averaging over the disorder, we obtain \cite{Levitov}
\begin{equation}
\Pi_{xx}(\mathbf{r}) =  \raisebox{-0.74\height}{\includegraphics[width=0.5\textwidth]{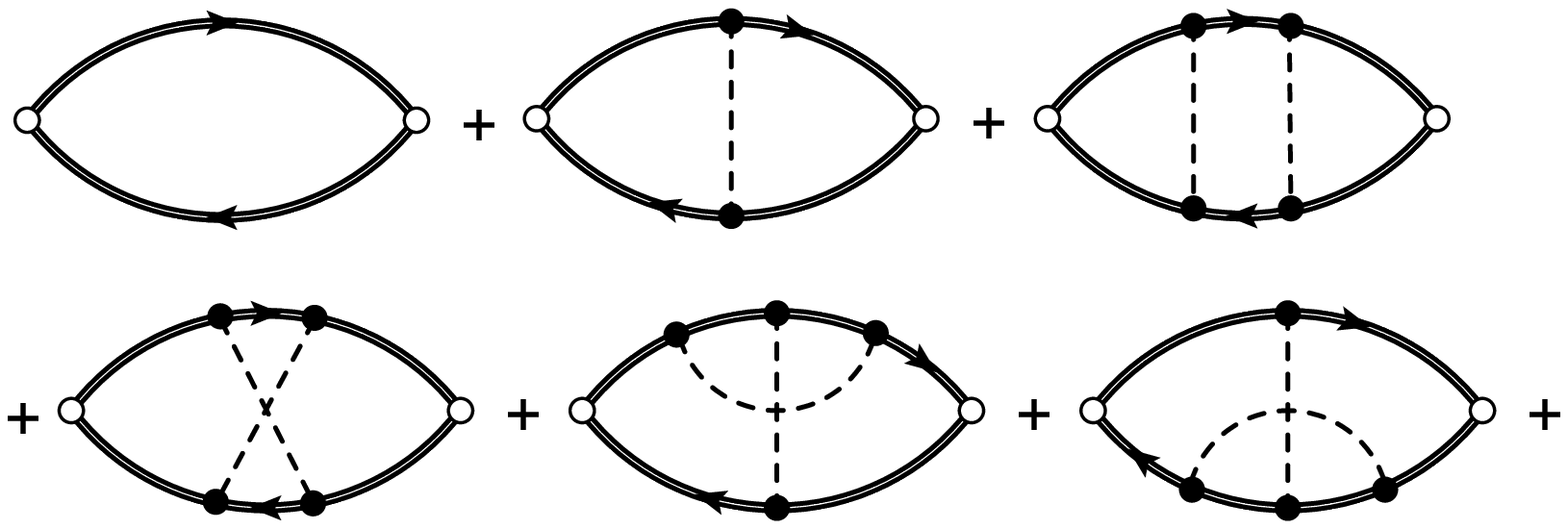}}
~ \raisebox{-0.65in}{$\cdots$}
\label{correlator-diagram-expansion}
\end{equation}
where the dashed line represents the impurity correlation function (\ref{imp-cor-func}) and the double
solid line represents the disorder-averaged quasiparticle propagators $\langle G\rangle_{\rm dis}$.
In the rainbow approximation, the latter is given by
\begin{equation}
\langle G\rangle_{\rm dis} \equiv \raisebox{-0.2\height}{\includegraphics[width=0.38\textwidth]{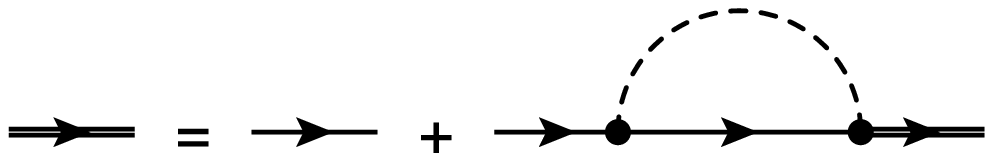}}  .
\label{diagram:propagator-average}
\end{equation}
It should be noted that, in the last equation, we ignored a correction to the self-energy linear in the
impurity potential. As we will show in Sec.~\ref{sec:width}, it has the form $\Sigma^{(1)} = n_{\rm imp}u_0$.
The effect of such a term is to shift the value of the chemical potential, $\mu^{*} = \mu-n_{\rm imp}u_0$.
Clearly, such a redefinition has no qualitative importance and, thus, will be ignored.

In three-dimensional systems, the contributions of the diagrams with crossing impurity lines, e.g., see
the second row of diagrams in Eq.~(\ref{correlator-diagram-expansion}), are usually suppressed 
compared to the diagrams without crossing lines \cite{Levitov}. Since in the present model there 
are no strict one-dimensional chiral fermions that are fully decoupled from the bulk states, its 
dynamics should be close to a three dimensional one. Therefore, one may expect that omitting 
the diagrams with crossing impurity lines provides a reasonable approximation in this case.

Then, the current-current correlation function takes the following diagrammatic form:
\begin{equation}
\Pi_{xx} = \vcenter{\hbox{\includegraphics[width=0.12\textwidth]{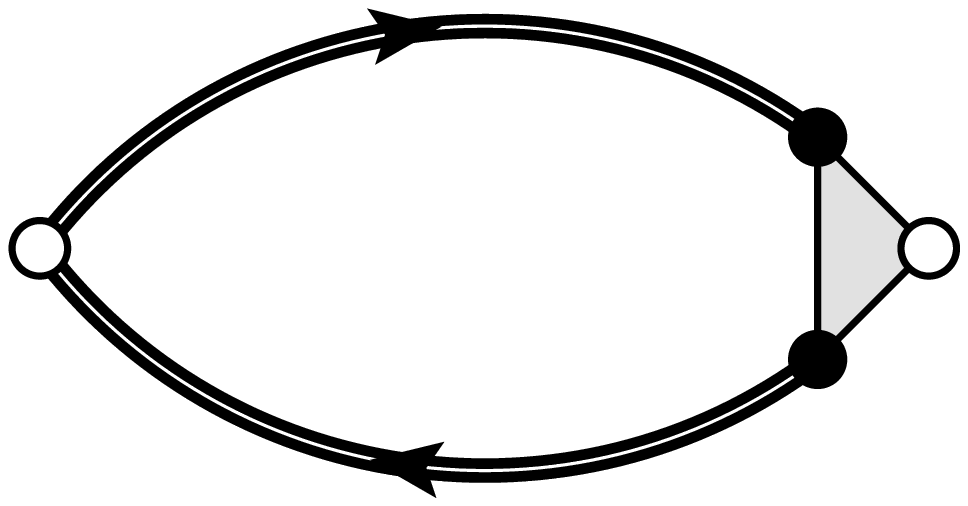}}}
=-e^2v_F^2\Omega\int\frac{d\omega}{2\pi} \frac{\partial n(\omega-\mu)}{\partial \omega}\int d^3\mathbf{r}^{\prime}d^3\mathbf{r}_1d^3\mathbf{r}_2 \mathrm{tr}\left[ \sigma_x
\langle G^A_s(\omega; \mathbf{r}, \mathbf{r}_1)\rangle_{\rm dis} \,
\Lambda_x\left(\omega; \mathbf{r}_{1}, \mathbf{r}^{\prime}, \mathbf{r}_{2}\right)
\langle G^R_s(\omega; \mathbf{r}_2, \mathbf{r})\rangle_{\rm dis} \right],
\label{vertex-correlator-1-vf-def}
\end{equation}
where we introduced the vertex function in the ladder approximation:
\begin{equation}
\Lambda_x\left(\omega; \mathbf{r}_{1}, \mathbf{r}^{\prime}, \mathbf{r}_{2}\right)  \equiv
\vcenter{\hbox{\includegraphics[width=0.2\textwidth]{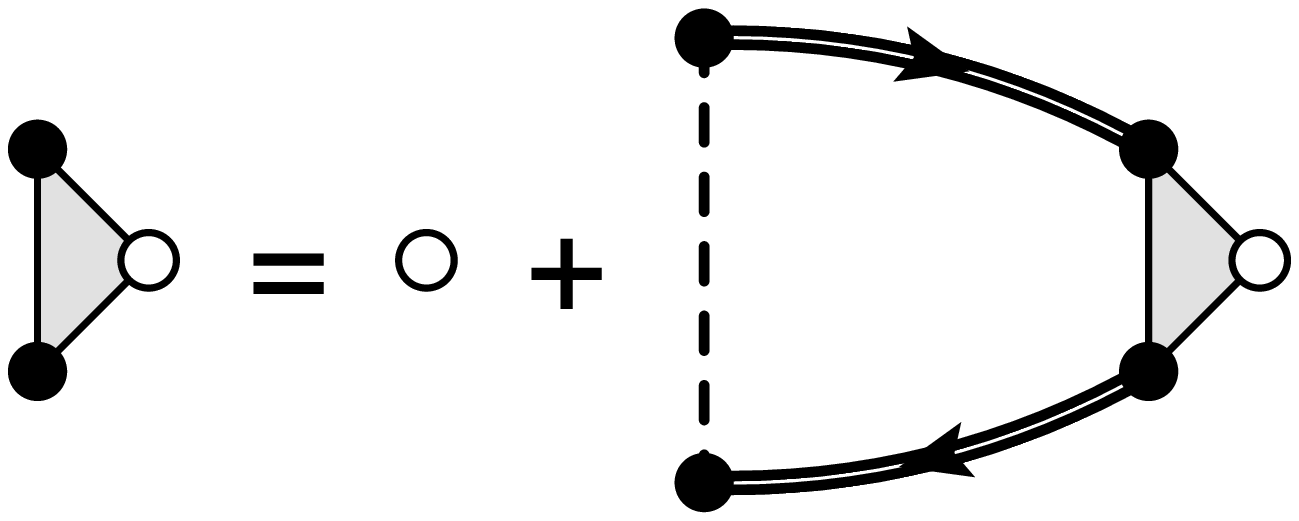}}}
\label{diagram:vertex}
\end{equation}
The corresponding explicit expression reads
\begin{equation}
\Lambda_x\left(\omega; \mathbf{r}_{1}, \mathbf{r}^{\prime}, \mathbf{r}_{2}\right) = \sigma_x \delta(\mathbf{r}_{1}-\mathbf{r}^{\prime})\delta(\mathbf{r}^{\prime}-\mathbf{r}_2) +D(\mathbf{r}_1, \mathbf{r}_2) \int d^3\mathbf{r}_1^{\prime} d^3\mathbf{r}_2^{\prime}
(-i)\langle G^A_s(\omega; \mathbf{r}_1, \mathbf{r}_1^{\prime})\rangle_{\rm dis}
\Lambda_x\left(\omega; \mathbf{r}_{1}^{\prime}, \mathbf{r}^{\prime}, \mathbf{r}_{2}^{\prime}\right)
(-i) \langle G^R_s(\omega; \mathbf{r}_2^{\prime}, \mathbf{r}_2)\rangle_{\rm dis}.
\label{vertex-def}
\end{equation}
By switching to a mixed coordinate and momentum space representation, we rewrite the expression for the
conductivity in the following form:
\begin{equation}
\sigma_{xx}(\mathbf{q}_{\parallel}, y) =  -\lim_{\Omega \to 0} \frac{i}{\Omega}\Pi_{xx}(\mathbf{q}_{\parallel}, y),
\label{vertex-conductivity-T0-mom}
\end{equation}
where the corresponding current-current correlator is given by
\begin{eqnarray}
\label{vertex-correlator-1-vf-def-mom}
\Pi_{xx}(\mathbf{q}_{\parallel}, y) &=& -e^2v_F^2\Omega\int\frac{d\omega}{2\pi} \frac{\partial n(\omega-\mu)}{\partial \omega}
\int \frac{d^2 \mathbf{k}_{\parallel}}{(2\pi)^2}  \int dy^{\prime} dy_1dy_2 \mathrm{tr}\Big[ \sigma_x
\langle G^A_s(\omega; \mathbf{k}_{\parallel}, y, y_1)\rangle_{\rm dis}
\nonumber\\
&\times&
\Lambda_x\left(\omega; \mathbf{k}_{\parallel}, \mathbf{k}_{\parallel}-\mathbf{q}_{\parallel},  y_{1}, y^{\prime}, y_{2}\right) \langle G^R_s(\omega;\mathbf{k}_{\parallel}-\mathbf{q}_{\parallel}, y_2, y)\rangle_{\rm dis} \Big].
\end{eqnarray}
In this study we are interested in the DC electrical conductivity. By taking into account that the corresponding
transport coefficient describes the linear response of the system to a spatially homogeneous electric field, we
set $\mathbf{q}_{\parallel}=0$. Then, the vertex function satisfies the following equation:
\begin{eqnarray}
\Lambda_x\left(\omega;\mathbf{k}_{\parallel},\mathbf{k}_{\parallel}, y_{1}, y^{\prime}, y_{2}\right) &=& \sigma_x\delta(y_{1}-y^{\prime})\delta(y^{\prime}-y_2) +\int \frac{d^2 \mathbf{l}_{\parallel}}{(2\pi)^2} \int dy_1^{\prime} dy_2^{\prime} D(\mathbf{k}_{\parallel} -\mathbf{l}_{\parallel}, y_1, y_2)\nonumber\\
&\times&(-i)\langle G^A_s(\mathbf{l}_{\parallel}, \omega; y_1, y_1^{\prime})\rangle_{\rm dis}  \Lambda_x\left(\omega;\mathbf{l}_{\parallel},\mathbf{l}_{\parallel}, y_{1}^{\prime}, y^{\prime}, y_{2}^{\prime}\right) (-i) \langle G^R_s(\omega;\mathbf{l}_{\parallel}, y_2^{\prime}, y_2)\rangle_{\rm dis}.
\label{vertex-def-mom}
\end{eqnarray}
In order to calculate the conductivity in this approximation, therefore, we would first need to calculate the
full quasiparticle propagator averaged over the disorder and then solve Eq.~(\ref{vertex-def-mom})
for the full vertex function. Finally, the knowledge of the propagator and vertex will allow us to derive the
expression for the correlator (\ref{vertex-correlator-1-vf-def-mom}) and, therefore, the conductivity
(\ref{vertex-conductivity-T0-mom}). Thus, in the next two subsections, we calculate both of them
using standard methods.

\subsection{Self-energy and full quasiparticle propagator}
\label{sec:width}

In this subsection we calculate the self-energy and determine the width of the Fermi arc states in the
full model (\ref{Hamiltonian-effective-M1}). To start with, let us remind that the surface and bulk Green's
functions in the absence of disorder are given by Eqs.~(\ref{S-surf}) and (\ref{S-bulk}), respectively.

Here we will determine the full Green's function for the surface states by calculating their self-energy to
the quadratic order in the disorder strength $u_0$. Formally, the definition for the self-energy is the same
as in Eq.~(\ref{disorder-SE}). However, one has to keep in mind that the internal quasiparticle propagator
in the corresponding diagram could be either the surface or bulk one. Therefore, there are different types
of contributions to the final result coming from the diagrams with the bulk or surface internal lines.

By expanding the full Green's function to the second order in powers of the disorder strength $u_0$,
see Eqs.~(\ref{diagram:propagator}) and (\ref{diagram:propagator-average}), we derive the following result:
\begin{equation}
(-i)G(\omega, \mathbf{k}_{\parallel}, \mathbf{k}_{\parallel}^{\prime}; y, y^{\prime}) =
S^{(0)}(\omega, \mathbf{k}_{\parallel}, \mathbf{k}_{\parallel}^{\prime}; y, y^{\prime})
+S^{(1)}(\omega, \mathbf{k}_{\parallel}, \mathbf{k}_{\parallel}^{\prime}; y, y^{\prime})
+S^{(2)}(\omega, \mathbf{k}_{\parallel}, \mathbf{k}_{\parallel}^{\prime}; y, y^{\prime}),
\label{S-series}
\end{equation}
where
\begin{equation}
S^{(0)}(\omega, \mathbf{k}_{\parallel}, \mathbf{k}_{\parallel}^{\prime}; y, y^{\prime}) =(2\pi)^2\delta(\mathbf{k}_{\parallel}- \mathbf{k}_{\parallel}^{\prime})(-i)S_{s}(\omega, \mathbf{k}_{\parallel}; y, y^{\prime}),
\label{S-0}
\end{equation}
and $S_{s}(\omega, \mathbf{k}_{\parallel}; y, y^{\prime})$ is the surface Green's function given by (\ref{S-surf}).
The first order correction is given by
\begin{eqnarray}
S^{(1)}(\omega, \mathbf{k}_{\parallel}, \mathbf{k}_{\parallel}^{\prime}; y, y^{\prime}) &=&\sum_j \int d y^{\prime \prime} (-i)S_{s}(\omega, \mathbf{k}_{\parallel}; y, y^{\prime \prime}) u_0 \delta(y^{\prime \prime}-y_j)e^{-i\mathbf{r}_{j}(\mathbf{k}_{\parallel}-\mathbf{k}_{\parallel}^{\prime})} (-i)S_{s}(\omega, \mathbf{k}_{\parallel}^{\prime}; y^{\prime \prime}, y^{\prime}) \nonumber \\
&=& \sum_j \int d y^{\prime \prime} \frac{p(k_z) (1+\sigma_x)e^{-(y+y^{\prime \prime})p(k_z)} }{\omega+i0+\mu-v_F k_x} u_0 \delta(y^{\prime \prime}-y_j)e^{-i\mathbf{r}_{j}(\mathbf{k}_{\parallel}-\mathbf{k}_{\parallel}^{\prime})}
\frac{p(k_z^{\prime})(1+\sigma_x)e^{-(y^{\prime \prime}+y^{\prime})p(k_z^{\prime})} }{\omega+i0+\mu-v_F k_x^{\prime}}.\nonumber \\
\label{S-1}
\end{eqnarray}
By averaging over the impurities using the prescription in Eq.~(\ref{disorder-average}) and integrating over
$y^{\prime \prime}$, we obtain
\begin{eqnarray}
\langle S^{(1)}(\omega, \mathbf{k}_{\parallel}, \mathbf{k}_{\parallel}^{\prime}; y, y^{\prime})\rangle_{\rm dis}
&=& \sum_j \frac{1}{V} u_0 (2\pi)^2 \delta(\mathbf{k}_{\parallel}-\mathbf{k}_{\parallel}^{\prime})
\frac{p(k_z)}{2} \frac{(1+\sigma_x)^2}{(\omega+i0+\mu-v_Fk_x)^2} e^{-(y+y^{\prime})p(k_z)} \nonumber \\
&=& n_{\rm imp}u_0 \frac{e^{(y+y^{\prime})p(k_z)} }{2p(k_z)} (2\pi)^2 \delta(\mathbf{k}_{\parallel}-\mathbf{k}_{\parallel}^{\prime})(-i)S_{s}(\omega, \mathbf{k}_{\parallel}; y, y^{\prime})(-i)S_{s}(\omega, \mathbf{k}_{\parallel}; y^{\prime}, y).
\label{S-1-a}
\end{eqnarray}
Let us note that, in order to extract the correct definition of the self-energy from this result,
it is instructive to compare the above expression with the formal expansion of the surface
Green's function to linear order in $\Sigma(k_z)$,
\begin{eqnarray}
\frac{p(k_z) (1+\sigma_x)e^{-(y+y^{\prime})p(k_z)}}{\omega +\mu -v_F k_x -\Sigma(k_z)} &=&
\frac{p(k_z) (1+\sigma_x)e^{-(y+y^{\prime})p(k_z)}}{\omega+\mu-v_F k_x}
+\frac{p(k_z) (1+\sigma_x)e^{-(y+y^{\prime})p(k_z)}}{(\omega+\mu-v_F k_x)^2} \Sigma(k_z)
\nonumber\\
&=& (-i)S_{s}(\omega, \mathbf{k}_{\parallel}; y^{\prime}, y)
+\frac{e^{(y+y^{\prime})p(k_z)} }{2p(k_z)}
(-i)S_{s}(\omega, \mathbf{k}_{\parallel}; y^{\prime}, y) \Sigma(k_z)
(-i)S_{s}(\omega, \mathbf{k}_{\parallel}; y^{\prime}, y) .
\label{Green-expansion-M}
\end{eqnarray}
Here we took into account that $(1+\sigma_x)^2=2(1+\sigma_x)$. By comparing this expansion
with the result in Eq.~(\ref{S-1-a}), we extract the following first-order correction to the real part
of self energy:
\begin{eqnarray}
\Sigma^{(1)}(\mathbf{k}_{\parallel}) = n_{\rm imp}u_0.
\label{SE-1}
\end{eqnarray}
This result is qualitatively the same as the leading-order correction to the self-energy in 
three-dimensional theories \cite{Levitov}. Its effect is to shift the chemical potential from $\mu$ to
$\mu^{*}$, where
\begin{equation}
\mu^{*}\equiv \mu-\Sigma^{(1)}(\mathbf{k}_{\parallel}) = \mu-n_{\rm imp}u_0.
\label{mu-eff}
\end{equation}
In the remainder of this paper, we will ignore this correction to the chemical potential
(or the difference between $\mu$ and $\mu^{*}$) due to its qualitative unimportance.

Next, we calculate the second-order correction
$S^{(2)}(\omega, \mathbf{k}_{\parallel}, \mathbf{k}_{\parallel}^{\prime}; y, y^{\prime})$
using the {\em surface} Green's function as an intermediate line in the second-order diagram in
Eq.~(\ref{diagram:propagator-average}), i.e.,
\begin{eqnarray}
S^{(2)}_s(\omega, \mathbf{k}_{\parallel}, \mathbf{k}_{\parallel}^{\prime}; y, y^{\prime})
&=&\sum_j \int d y_1 dy_2 \int \frac{d^2 \mathbf{q}_{\parallel}}{(2\pi)^2}
(-i)S(\omega, \mathbf{k}_{\parallel}; y, y_1) u_0 \delta(y_1-y_j)
e^{-i\mathbf{r}_{j}(\mathbf{k}_{\parallel}-\mathbf{q}_{\parallel})} (-i)S(\omega, \mathbf{q}_{\parallel}; y_1, y_2)
\nonumber \\
&\times& u_0 \delta(y_2-y_j)e^{-i\mathbf{r}_{j}(\mathbf{q}_{\parallel}-\mathbf{k}_{\parallel}^{\prime})}(-i)S(\omega, \mathbf{k}_{\parallel}^{\prime}; y_2, y^{\prime})
= \sum_j u_0^2e^{-i\mathbf{r}_{j}(\mathbf{k}_{\parallel}-\mathbf{k}_{\parallel}^{\prime})}  e^{-p(k_z)(y+y^{\prime})}
\nonumber \\
&\times& \int \frac{d^2 \mathbf{q}_{\parallel}}{(2\pi)^2} e^{-2p(k_z)y_j-2p(q_z)y_j}
\frac{(1+\sigma_x)^3 [p(k_z)]^2 p(q_z)}{(\omega+i0+\mu-v_F k_x)^2(\omega+i0+\mu-v_F q_x)}.
\label{S-2}
\end{eqnarray}
After the averaging over impurities, we then obtain
\begin{eqnarray}
\langle S^{(2)}_s(\omega, \mathbf{k}_{\parallel}, \mathbf{k}_{\parallel}^{\prime}; y, y^{\prime})\rangle_{\rm dis}
&=& \frac{N_{\rm imp}}{V} u_0^2 \delta(\mathbf{k}_{\parallel}-\mathbf{k}_{\parallel}^{\prime}) \int d^2 \mathbf{q}_{\parallel}
\frac{(1+\sigma_x)^3e^{-p(k_z)(y+y^{\prime})}  }{(\omega+i0+\mu-v_Fk_x)^2(\omega+i0+\mu-v_Fq_x)}
\frac{[p(k_z)]^2p(q_z)}{2p(k_z)+2p(q_z)}
\nonumber \\
&=& \frac{-i\pi}{v_F} n_{\rm imp}u_0^2 \delta(\mathbf{k}_{\parallel}-\mathbf{k}_{\parallel}^{\prime})
\frac{(1+\sigma_x)^3 e^{-p(k_z)(y+y^{\prime})}}{(\omega+i0+\mu-v_Fk_x)^2} \frac{[p(k_z)]^2}{2}
\int_{-\sqrt{m}}^{\sqrt{m}} dq_z \left[1- \frac{p(k_z)}{p(k_z)+p(q_z)}\right]. \nonumber \\
\label{S-2-a}
\end{eqnarray}
Here we took into account the retarded nature of the Green's functions and used the finiteness
of the $z$ component of momentum. The latter condition stems from the normalizability of the
wave function. Furthermore, the result in Eq.~(\ref{S-2-a}) can be rewritten as follows:
\begin{eqnarray}
\langle S^{(2)}_s(\omega, \mathbf{k}_{\parallel}, \mathbf{k}_{\parallel}^{\prime}; y, y^{\prime})\rangle_{\rm dis}
&=& \frac{-i}{4\pi v_F}  n_{\rm imp}u_0^2 (2\pi)^2\delta(\mathbf{k}_{\parallel}-\mathbf{k}_{\parallel}^{\prime})e^{p(k_z)(y+y^{\prime})}
(-i)S(\omega, \mathbf{k}_{\parallel}; y, y^{\prime})(-i)S(\omega, \mathbf{k}_{\parallel}; y^{\prime}, y)\nonumber \\
&\times& \left[2\sqrt{m}+ \frac{2v_Fp(k_z)}{\sqrt{-\gamma v_F p(k_z)-m\gamma^2}} \arctan{\left(\frac{\sqrt{\gamma m}}
{\sqrt{-v_Fp(k_z)-\gamma m}}\right)}\right].
\label{S-2-b}
\end{eqnarray}
Thus, we finally extract the second-order correction to the self-energy
\begin{eqnarray}
\Sigma^{(2)}_s(\mathbf{k}_{\parallel}) = -i\frac{ \gamma(m-k_z^2)}{\pi v_F^2} n_{\rm imp}u_0^2\left[\sqrt{m} -\frac{(k_z^2-m)}
{\sqrt{k_z^2-2m}} \arctan{\left(\frac{\sqrt{m}}{\sqrt{k_z^2-2m}}\right)}\right].
\label{SE-2}
\end{eqnarray}
As we see, this correction is imaginary and has a nontrivial dependence on the wave vector $k_z$.
This means that the surface quasiparticles receive a nonzero width contribution, given by
\begin{equation}
\Gamma_s(k_z)=\frac{\gamma(m-k_z^2)}{\pi v_F^2} n_{\rm imp}u_0^2 \left[\sqrt{m} -\frac{(k_z^2-m)}{\sqrt{k_z^2-2m}}
\arctan{\left(\frac{\sqrt{m}}{\sqrt{k_z^2-2m}}\right)}\right].
\label{Gamma-FA}
\end{equation}
We note, however, that the above result is not complete yet. To the same second order in the
disorder strength, there is also another contribution, in which the intermediate line in the
second-order diagram in Eq.~(\ref{diagram:propagator-average}) is given by the {\em bulk} Green's
function (\ref{S-bulk}), i.e.,
\begin{eqnarray}
\langle S^{(2)}_b(\omega, \mathbf{k}_{\parallel}, \mathbf{k}_{\parallel}^{\prime}; y, y^{\prime})\rangle_{\rm dis}
&=& \Big\langle \sum_j \int d y_1 dy_2 \int \frac{d^2 \mathbf{q}_{\parallel}}{(2\pi)^2} (-i)S(\omega, \mathbf{k}_{\parallel}; y, y_1)
u_0 \delta(y_1-y_j)e^{-i\mathbf{r}_{j}(\mathbf{k}_{\parallel}-\mathbf{q}_{\parallel})}
\nonumber \\
&\times& (-i)S_b(\omega, \mathbf{q}_{\parallel}; y_1-y_2)  u_0 \delta(y_2-y_j)e^{-i\mathbf{r}_{j}(\mathbf{q}_{\parallel}
-\mathbf{k}_{\parallel}^{\prime})}(-i)S(\omega, \mathbf{k}_{\parallel}^{\prime}; y_2, y^{\prime})\Big\rangle_{\rm dis}.
\label{S-2-bulk}
\end{eqnarray}
After simplification, this result can be rewritten equivalently as follows:
\begin{eqnarray}
\langle S^{(2)}_b(\omega, \mathbf{k}_{\parallel}, \mathbf{k}_{\parallel}^{\prime}; y, y^{\prime})\rangle_{\rm dis}  &=&  n_{\rm imp}u_0^2 (2\pi)^2\delta(\mathbf{k}_{\parallel}-\mathbf{k}_{\parallel}^{\prime})e^{p(k_z)(y+y^{\prime})}  (-i)S(\omega, \mathbf{k}_{\parallel}; y, y^{\prime})(-i)S(\omega, \mathbf{k}_{\parallel}; y^{\prime}, y) \frac{e^{(y+y^{\prime})p(k_z)}}{2p(k_z)} \nonumber\\
&\times& \int \frac{d^3 \mathbf{q}}{(2\pi)^3} \frac{\left[\omega+\mu +v_Fq_x\sigma_x\right]}{(\omega+\mu )^2-\gamma^2\left(q_z^2-m\right)^2-v_F^2q_{\parallel}^2+i0\sign{(\omega+\mu )}}.
\label{S-2-b-bulk}
\end{eqnarray}
While the general result is rather complicated, we can extract its imaginary part
analytically by using the Sokhotski's formula. After integrating over $\mathbf{q}$,
we derive the following result:
\begin{equation}
\mathrm{Im}\left[\Sigma^{(2)}_b(\omega)\right]  =
-n_{\rm imp} u_0^2  \frac{|\omega+\mu | }{4 v_F^2 \pi}
\left[\sqrt{m+\frac{|\omega+\mu |}{\gamma}}-\theta\left(m-\frac{|\omega+\mu |}{\gamma}\right)
\sqrt{m-\frac{|\omega+\mu |}{\gamma}} \right],
\label{S-2bb}
\end{equation}
which has a nontrivial dependence on the energy $\omega$.
The corresponding contribution to the quasiparticle width is given by
\begin{equation}
\Gamma_{b}(\omega)\equiv n_{\rm imp} u_0^2  \frac{|\omega+\mu |}{4 v_F^2 \pi} \left[\sqrt{m+\frac{|\omega+\mu |}{\gamma}}-\theta\left(m-\frac{|\omega+\mu |}{\gamma}\right)
\sqrt{m-\frac{|\omega+\mu |}{\gamma}} \right].
\label{Gamma-FB}
\end{equation}
Finally, by combining the results in Eqs.~(\ref{Gamma-FA}) and (\ref{Gamma-FB}), we derive the
total quasiparticle width of surface quasiparticles, i.e.,
\begin{equation}
\Gamma(\omega; k_z)=\Gamma_{s}(k_z)+\Gamma_{b}(\omega).
\label{Gamma-FA-tot}
\end{equation}
The corresponding full surface Green's function, averaged over disorder, is given by
\begin{equation}
\langle G(\omega, \mathbf{k}_{\parallel}; y, y^{\prime}) \rangle_{\rm dis} \simeq
\frac{ 2 ip(k_z) e^{-(y+y^{\prime})p(k_z)}}{\omega+\mu-v_Fk_x+i\Gamma(\omega, k_z)} \frac{1+\sigma_x}{2}.
\label{S-wave-func-FA}
\end{equation}

\subsection{Vertex correction}
\label{sec:vertex}

In this subsection we calculate the full vertex function in the ladder approximation.
The explicit form of the equation for the vertex (\ref{vertex-def-mom}) reads
\begin{eqnarray}
\Lambda_x\left(\omega;\mathbf{k}_{\parallel},\mathbf{k}_{\parallel}, y_{1}, y^{\prime}, y_{2}\right)
&=& \sigma_x\delta(y_{1}-y_2)\delta(y^{\prime}-y_1) + n_{\rm imp}u_0^2\delta(y_1-y_2)
\int \frac{d^2 \mathbf{l}_{\parallel}}{(2\pi)^2}\int dy_1^{\prime} dy_2^{\prime}
\frac{(1+\sigma_x)p(l_z)e^{-(y_1+y_1^{\prime})p(l_z)}}{\left[\omega-v_Fl_x-i\Gamma(\omega, l_z)\right]} \nonumber \\
&\times&\Lambda_x(\omega, \mathbf{l}_{\parallel},\mathbf{l}_{\parallel}; y_1^{\prime}, y^{\prime},y_2^{\prime})
\frac{(1+\sigma_x)p(l_z)e^{-(y_2+y_2^{\prime})p(l_z)}}{\left[\omega-v_Fl_x+i\Gamma(\omega, l_z)\right]}.
\label{vertex-01}
\end{eqnarray}
The matrix structure on the right-hand side of the above equation suggests that
the solution for the vertex function should take the following form:
\begin{equation}
\Lambda_x(\omega, \mathbf{k}_{\parallel},\mathbf{k}_{\parallel}; y_{1}, y^{\prime}, y_{2})
= \sigma_x\delta(y_{1}-y_2)\delta(y^{\prime}-y_1)
+ \delta(y_{1}-y_{2})\frac{1+\sigma_x}{2} f(\omega, y_1,y^{\prime}).
\label{vertex-ansatz}
\end{equation}
By substituting such an ansatz into Eq.~(\ref{vertex-01}), we find that the unknown
function $f(\omega, y_1,y^{\prime})$ should satisfy the following equation:
\begin{equation}
f(\omega, y_1, y^{\prime}) = F(\omega, y_1, y^{\prime})+\int dy_1^{\prime}
F(\omega, y_1,y_1^{\prime}) f(\omega, y_1^{\prime}, y^{\prime}),
\label{vertex-sys}
\end{equation}
where
\begin{eqnarray}
F(\omega, y_1, y^{\prime})\equiv \frac{2n_{\rm imp}u_0^2}{v_F} \int_{-\sqrt{m}}^{\sqrt{m}}
\frac{d l_z}{2\pi} \frac{p^2(l_z)e^{-2(y_1+y^{\prime})p(l_z)}}{\Gamma(\omega, l_z)}.
\label{vertex-F}
\end{eqnarray}
By compiling this function numerically, we found that it can be approximated quite well by the following
fit:
\begin{eqnarray}
F(\omega, y_1, y^{\prime})\approx \frac{a(\omega)}{1+d(\omega)(y_1+y^{\prime})+b(\omega)(y_1+y^{\prime})^2},
\label{vertex-F-fit-2}
\end{eqnarray}
where the three fitting parameters are functions of $\omega$. At $\omega=0$, for example,
their numerical values are $a(0)=0.106~\mbox{\AA}^{-1}$ and $b(0)=0.004~\mbox{\AA}^{-2}$,
and $d(0)=0.032~\mbox{\AA}^{-1}$.

Equation~(\ref{vertex-sys}) is a Fredholm integral equation of the second kind. When the norm of the kernel
is sufficiently small, i.e.,
\begin{equation}
N_F(\omega)\equiv\left( \int_0^{\infty} dy \int_0^{\infty} dy^{\prime} F^2\left(\omega, y, y^{\prime}\right) \right)^{1/2}<1,
\label{vertex-norm}
\end{equation}
its solution can be obtained by iterations. The formal result in this case reads
\begin{equation}
f(\omega, y, y^{\prime})=F(\omega, y, y^{\prime})+\sum_{n=1}^{\infty}
\int_0^{\infty} dy^{\prime\prime} K_n\left(\omega, y, y^{\prime\prime}\right)
F(\omega, y^{\prime\prime},y^{\prime}),
\label{vertex-Fredholm-sol}
\end{equation}
where, by definition, $K_1(\omega, y, y^{\prime}) \equiv F(\omega, y, y^{\prime})$ and
\begin{equation}
K_n\left(\omega, y, y^{\prime}\right) = \int_0^{\infty} d y^{\prime\prime} F\left(\omega, y, y^{\prime\prime}\right)
K_{n-1}\left(\omega, y^{\prime\prime},y^{\prime}\right).
\label{vertex-Fredholm-sol-2}
\end{equation}
This solution defines a convergent series at sufficiently large values of $\omega$ [e.g., for
$\omega \approx \gamma m/2$, $N_F(\omega)\approx0.8$]. For small $\omega$ (i.e.,
$\omega \lesssim 0.04\gamma m$), an alternative algorithm for solving the corresponding
Fredholm equation should be utilized.

\subsection{Fermi arc conductivity}
\label{sec:Arc-Conductivity-Full}

By combining the results of the previous two subsections, in this subsection we
calculate the conductivity of the Fermi arc states.

We start by presenting the mixed coordinate-momentum space representation of
the spectral function for the surface Fermi arc states,
\begin{eqnarray}
A(\omega, \mathbf{k}_{\parallel}; y, y^{\prime}) &=&
\frac{i}{2\pi} \left[\langle G^{R}(\omega, \mathbf{k}_{\parallel}; y, y^{\prime})
\rangle_{\rm dis}- \langle G^{A}(\omega,\mathbf{k}_{\parallel}; y, y^{\prime})\rangle_{\rm dis}\right]_{\mu=0}
 \nonumber \\
&=& \frac{i}{\pi} (1+\sigma_x)p(k_z)e^{-(y+y^{\prime})p(k_z)}\frac{\Gamma(\omega, k_z)}{(\omega-
v_Fk_x)^2+\Gamma^2(\omega, k_z)}\theta(m-k_z^2),
\label{spectral-function-def-FA}
\end{eqnarray}
where $\Gamma(\omega,k_z)$ is given by Eq.~(\ref{Gamma-FA-tot}). The corresponding Green's function
can be restored from this spectral function by using the relation in Eq.~(\ref{Green-spectral-def}).

The $xx$ component of the dc conductivity tensor is defined by Eq.~(\ref{vertex-conductivity-T0-mom})
evaluated at $\mathbf{q}_{\parallel}=0$. By making use of the Green's function representation in terms
of the spectral function (\ref{spectral-function-def-FA}), we rewrite the expression for the conductivity
as follows:
\begin{eqnarray}
\sigma_{xx}^{\rm tot}(y) &=& -\lim_{\Omega\to0}\frac{i}{\Omega} e^2v_F^2 T
\sum_{l=-\infty}^{\infty} \int\frac{d^2 \mathbf{k}_{\parallel}}{(2\pi)^2}\int\int\int dy^{\prime} dy_1 dy_2
\int \int d\omega d\omega^{\prime} \nonumber\\
&\times& \frac{\mathrm{tr}\left[\sigma_x A(\omega, \mathbf{k}_{\parallel}; y, y_1)
\Lambda_x\left(\omega,\omega^{\prime};\mathbf{k}_{\parallel},\mathbf{k}_{\parallel}, y_{1}, y^{\prime}, y_{2}\right)
A(\omega^{\prime}, \mathbf{k}_{\parallel}; y_2, y)\right]}{\left(i\omega_l+\mu-\omega\right)\left(i\omega_l
- \Omega-i0+\mu-\omega^{\prime}\right)},
\label{conductivity-calc-1-tot}
\end{eqnarray}
where the vertex for $\omega=\omega^{\prime}$ is given by Eq.~(\ref{vertex-def-mom}).
We emphasize that the knowledge of the vertex correction at $\omega=\omega^{\prime}$ is
sufficient for calculating the real part of the dc conductivity.

To start with, let us first calculate the real part of the $xx$ component of the conductivity tensor
(\ref{conductivity-calc-1-tot}) without the vertex correction. In other words, we use the bare
vertex, $\Lambda^{(0)}_x\left(\omega;\mathbf{k}_{\parallel},\mathbf{k}_{\parallel}, y_{1}, y^{\prime}, y_{2}\right) = \sigma_x \delta(y^{\prime}-y_1)\delta(y^{\prime}-y_2)$.
The corresponding result reads
\begin{eqnarray}
\mathfrak{Re} \sigma_{xx}^{(0)}(y) &=&
e^2v_F^2 \int_0^{\infty} dy^{\prime} \int_{-\sqrt{m}}^{\sqrt{m}} \frac{dk_z}{2\pi}
\int \frac{dk_x d\omega}{(2\pi)^2} \frac{4p^2(k_z)}{4T\cosh^2{\left(\frac{\omega-\mu}{2T}\right)}}
\frac{e^{-2(y+y^{\prime})p(k_z)}}{(\omega-v_Fk_x)^2+\Gamma^2(\omega, k_z)} \nonumber\\
&=& \frac{e^2v_F}{4\pi^2} \int_{-\sqrt{m}}^{\sqrt{m}} dk_z \int d\omega
\frac{p(k_z)}{4T\cosh^2{\left(\frac{\omega-\mu}{2T}\right)}}\frac{e^{-2(y+y^{\prime})p(k_z)}}{\Gamma(\omega, k_z)}.
\label{conductivity-calc-Zero}
\end{eqnarray}
The first-order correction to this result is given by
\begin{eqnarray}
\mathfrak{Re} \sigma_{xx}^{(1)}(y) &=&  e^2v_F^2 \int \frac{d\omega}{2\pi} \frac{1}{4T\cosh^2{\left(\frac{\omega-\mu}{2T}\right)}} \int dy^{\prime} dy_1dy_2 \int \frac{d^2 \mathbf{k}_{\parallel}}{(2\pi)^2} \mathrm{tr}\Bigg[ \sigma_x \frac{p(k_z)(1+\sigma_x)e^{-(y+y_1)p(k_z)} \Gamma(\omega, k_z)}{(\omega-v_Fk_x)^2+\Gamma^2(\omega, k_z)} \nonumber \\
&\times&\delta(y_1-y_2) n_{\rm imp}u_0^2 \int \frac{d^2 \mathbf{l}_{\parallel}}{(2\pi)^2} \frac{(1+\sigma_x)^2\sigma_x p^2(l_z)e^{-2(y^{\prime}+y_1)p(l_z)}}{(\omega-v_Fl_x)^2+\Gamma^2(\omega, l_z)} \frac{p(k_z)(1+\sigma_x)e^{-(y_2+y)p(k_z)} \Gamma(\omega, k_z)}{(\omega-v_Fk_x)^2+\Gamma^2(\omega, k_z)}  \Bigg] \nonumber\\
&=& \frac{e^2}{16\pi^3} n_{\rm imp}u_0^2 \int d\omega \frac{1}{4T \cosh^2{\left(\frac{\omega-\mu}{2T}\right)}} \int_{-\sqrt{m}}^{\sqrt{m}} \int_{-\sqrt{m}}^{\sqrt{m}} \frac{dk_zdl_z}{\Gamma(\omega, k_z)\Gamma(\omega, l_z)} \frac{e^{-2yp(k_z)} p(l_z)p^2(k_z)}{\left[p(k_z)+p(l_z)\right]}.
\label{conductivity-calc-First}
\end{eqnarray}
Here we took into account the leading-order correction to the vertex with a single impurity correlator
bridging the incoming fermion lines in the diagram.

\begin{figure*}[!ht]
\begin{center}
\includegraphics[width=0.32\textwidth]{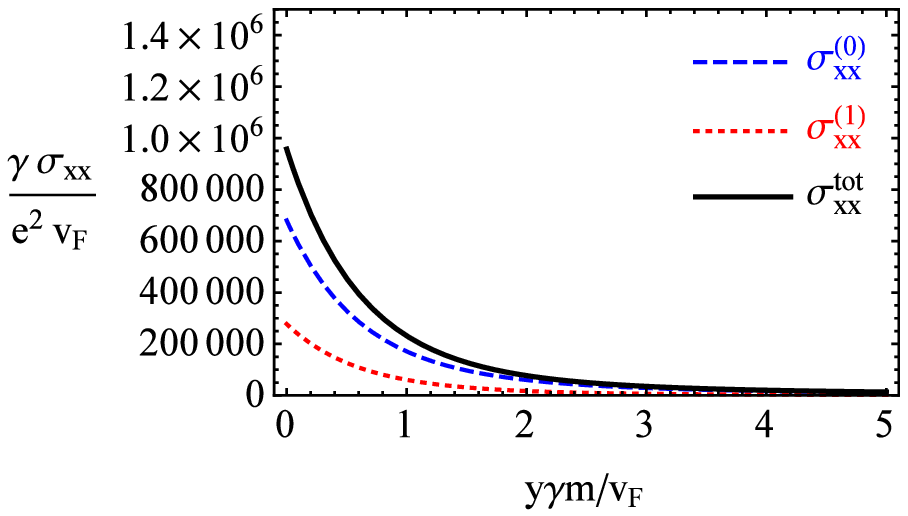} \hfill
\includegraphics[width=0.32\textwidth]{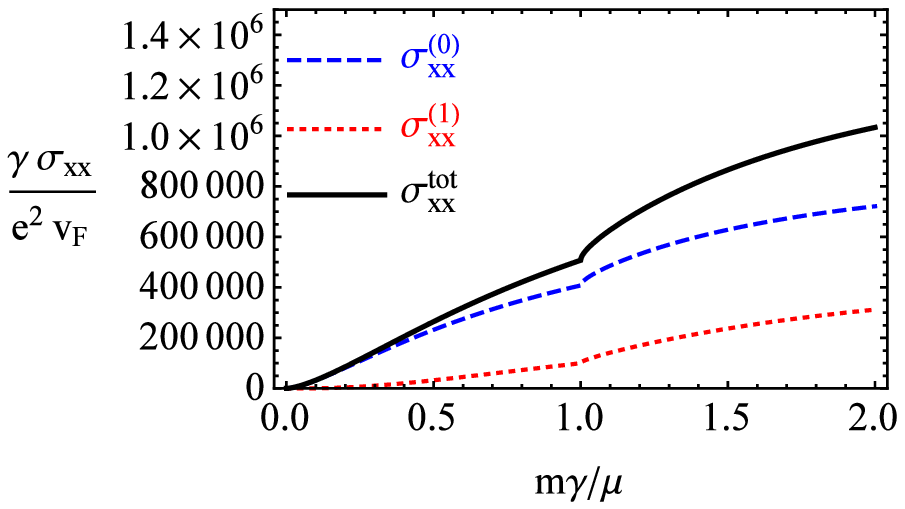} \hfill
\includegraphics[width=0.32\textwidth]{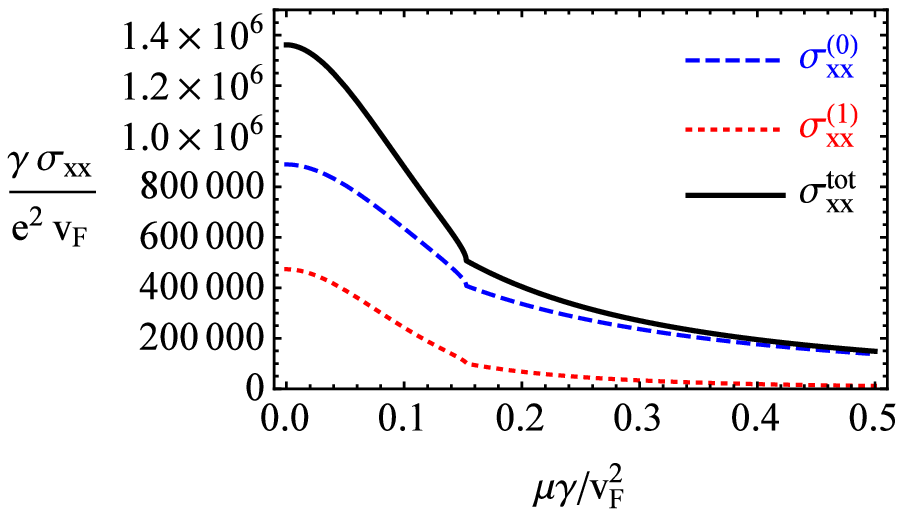}
\end{center}
\caption{(Color online) The real part of the surface conductivity at $T=0$ as a function
of distance from the surface $y$ (left panel), the parameter $m$
(middle panel), and the chemical potential $\mu$ (right panel). In the left and middle panels
$\mu=0.05~\mbox{eV}$, in the middle and right panels $y=0$. The other model parameters are
$u_0=0.1~\mbox{eV\, \AA}^{3}$ and $n_{\rm imp}=10^{-3}~\mbox{\AA}^{-3}$.}
\label{fig:sigma-xx-y-tot}
\end{figure*}

Our numerical results for the zero-temperature conductivity of the surface Fermi arc states
are summarized in Fig.~\ref{fig:sigma-xx-y-tot}. In each panel, we show (i) the leading-order
result $\mathfrak{Re} \sigma_{xx}^{(0)}(y)$ without any vertex corrections (blue dashed lines),
(ii) the first-order correction $\mathfrak{Re} \sigma_{xx}^{(1)}(y)$ (red dotted lines), as well as
(iii) the total conductivity $\mathfrak{Re} \sigma_{xx}^{\rm tot}(y)\equiv \mathfrak{Re} \sigma_{xx}^{(0)}(y)
+\mathfrak{Re} \sigma_{xx}^{(1)}(y)$ (black solid lines). For the strength of the disorder
potential and the density of impurities, we used the values $u_0=0.1~\mbox{eV\, \AA}^{3}$ and
$n_{\rm imp} = 10^{-3}~\mbox{\AA}^{-3}$, respectively.

The left panel in Fig.~\ref{fig:sigma-xx-y-tot} shows the real part of the conductivity as a function
of the distance $y$, measured from the surface into the bulk. The result is shown for a
fixed value of the chemical potential, $\mu=0.05~\mbox{eV}$. This choice of $\mu$ is sufficiently large
to be still marginally in the regime of reliable perturbative calculations, where the vertex correction is not
too large. This is also confirmed by the fact that the leading-order result dominates the total conductivity.
From the functional dependence we see that, as expected, the Fermi arc conductivity takes the highest
value on the surface ($y=0$) and rapidly decreases as the function of $y$.

The dependence of the conductivity on the parameter  
$\gamma m/\mu$ is shown in the middle panel of Fig.~\ref{fig:sigma-xx-y-tot}. The results are
plotted for $y=0$ and the same fixed value of the chemical potential, $\mu=0.05~\mbox{eV}$.
Despite the fact that the quasiparticle width (\ref{Gamma-FA}) has a very nontrivially dependence
on the parameter $m$, the conductivity appears to always grow with $m$. It is also
interesting to note that the slopes of the functional dependence are different at small and large
values of $\gamma m/\mu$, showing a characteristic (van Hove) kink at $m=\mu/\gamma$.
The existence of such a kink can be explained by the Lifshitz transition \cite{Lifshitz}.
Indeed, as one can see from the left panel of Fig.~\ref{fig:Fermi-surf}, two disjoined sheets
of the Fermi surface  merge into a single one at $m=\mu/\gamma$ resulting in the van Hove kink.
Note that the corresponding phenomenon was observed in three-dimensional Dirac semimetal
$\mathrm{Na_3Bi}$ \cite{Xu:Lifshitz}.

The right panel in Fig.~\ref{fig:sigma-xx-y-tot} shows the conductivity as a function of the 
chemical potential $\mu$ at fixed position $y=0$ (on the surface). One of the main features of the
corresponding result is that the total surface conductivity decreases with $\mu$. At first sight, this
may appear counterintuitive. However, this is indeed the case and the explanation is quite simple.
Because of the specific mechanism of dissipation, i.e., scattering of the surface Fermi arc
states into the bulk, the width of surface quasiparticles grows when the density of bulk states
increases. Indeed, such an increase expands the available phase space for scattered states
into the bulk and, thus, damps the surface conductivity.

From the functional dependence of the conductivity on the chemical potential shown in
the right panel of Fig.~\ref{fig:sigma-xx-y-tot}, we observe another interesting detail. In
agreement with our analytical estimates in Sec.~\ref{sec:vertex}, the validity of the
perturbative regime is limited only to the case of sufficiently large energies, or chemical
potentials, $\mu\gamma/v_F^2\gtrsim 0.05$, where the vertex corrections are under
control. For smaller values of $\mu$, a nonperturbative solution of the integral equation
for the vertex function (\ref{vertex-sys}) is needed. This is also supported by the
calculated dependence of the surface conductivity on $\mu$, which reveals that the
leading-order conductivity and the first correction become comparable at
$\mu\gamma/v_F^2\ll 0.05$.

\begin{figure*}[!ht]
\begin{center}
\includegraphics[width=0.32\textwidth]{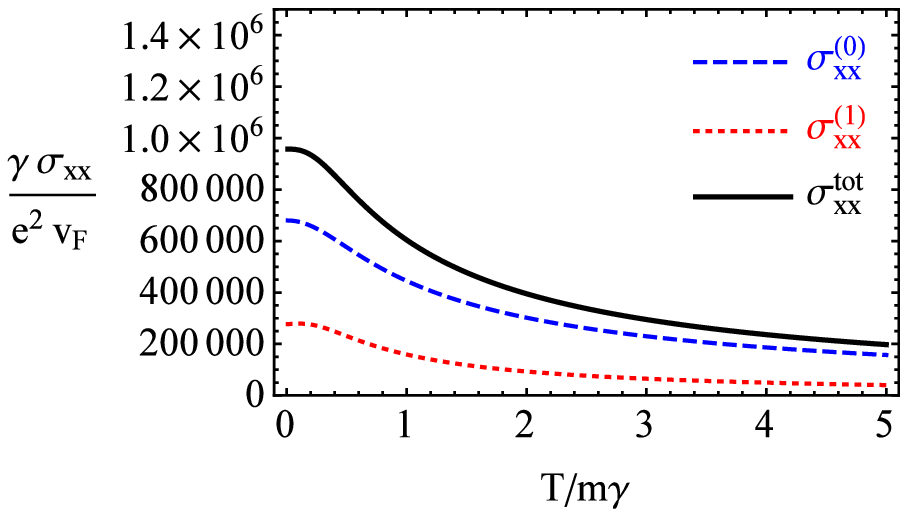}\hfill
\includegraphics[width=0.32\textwidth]{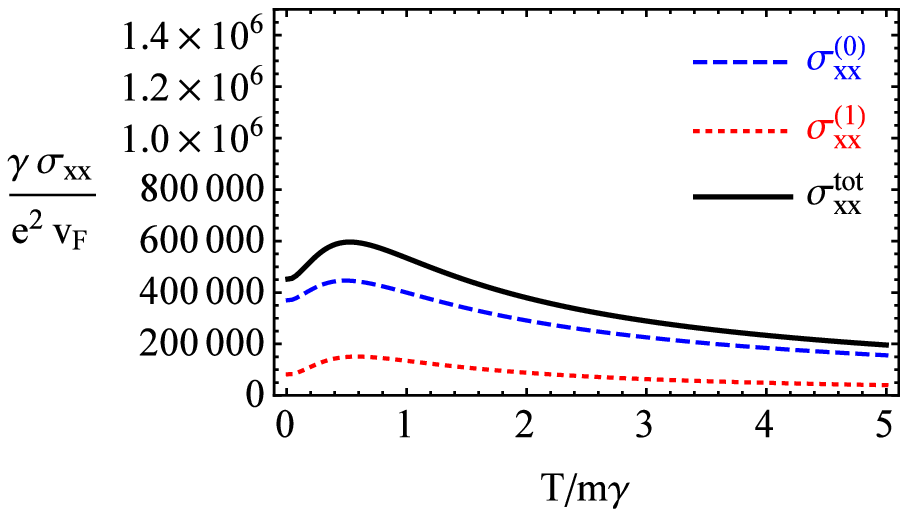}\hfill
\includegraphics[width=0.32\textwidth]{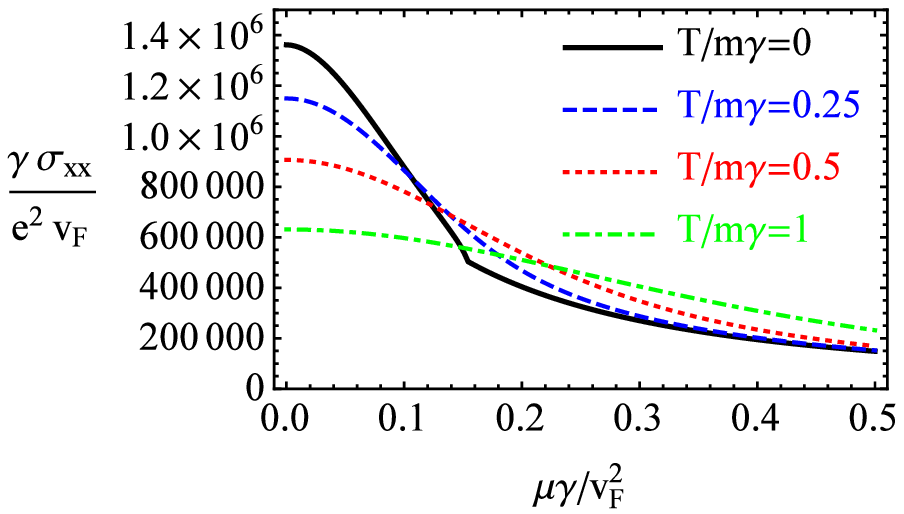}
\end{center}
\caption{(Color online) The real part of the surface conductivity as a function of temperature
$T$ at $\mu=0.05~\mbox{eV}$ (left panel), $\mu=0.1~\mbox{eV}$ (middle panel), as well as
as a function of the chemical potential $\mu$ at several different temperatures (right panel).
The other model parameters are $u_0=0.1~\mbox{eV\, \AA}^{3}$ and $n_{\rm imp}=10^{-3}~\mbox{\AA}^{-3}$.}
\label{fig:sigma-xx-T-tot-mu}
\end{figure*}

The finite temperature results for the conductivity due to surface Fermi arc states are
shown in Fig.~\ref{fig:sigma-xx-T-tot-mu} for $y=0$
and several representative values of parameters. By comparing the temperature
dependence of the conductivity at different values of the chemical potentials, we find
that there are two qualitatively different dynamical regimes. At sufficiently small values
of $\mu$ (i.e., $\mu \gamma/v_F^2\lesssim 0.06$), the conductivity monotonically
decreases with temperature. At larger values, there is a nonmonotonic dependence with
a local maximum at a temperature that scales approximately as the chemical potential,
\begin{equation}
\frac{T_{\rm peak}}{\gamma m}\approx \left(-0.2+4.1 \frac{\mu \gamma}{v_F^2}\right)
\theta\left(\frac{\mu \gamma}{v_F^2}-0.06\right).
\label{conductivity-T-max-vert-1}
\end{equation}
The underlying reason for the nonmonotonic behavior of the conductivity and the
location of the maximum around $T_{\rm peak}$ are not completely transparent
from the details of our analysis. It is natural to expect, however, that the roots of
this dependence lie in the underlying mechanism of the dissipation of the surface
states.

\section{Summary and discussions}
\label{sec:Conclusion}

In this paper, we addressed the question whether the charge transport of the Fermi arcs in Weyl semimetals
is dissipative or non-dissipative. Because of the topologically protected nature of the surface Fermi arc states,
one would naively expect that the Fermi arc transport should be non-dissipative. We found, however, that the
transport is, in fact, dissipative. The fundamental physical reason for the dissipation is the presence of gapless
bulk states in Weyl semimetals, whose low-energy dynamics is not fully decoupled from the surface Fermi
arc states. Indeed, by making use of the simplest model of a short-ranged quenched disorder, we showed that
the Fermi arc quasiparticles can scatter into the bulk states in addition to scattering into other surface Fermi
arc states. This nondecoupling implies that there is no well-defined effective theory of Fermi 
arcs in Weyl semimetals in the presence of quenched disorder. The scattering of quasiparticles
from the surface into the bulk and vice versa leads to the dephasing of the Fermi arc states and, thus,
dissipation. This is in contrast to the case of topological insulators, in which bulk states are gapped. It
appears that our findings agree with the generic conclusions in Refs.~\cite{QPI:Mitchell, QPI:Derry}, claiming
that an effective theory of surface states in gapless systems can not be formulated.

We calculated the dc conductivity of the Fermi arc states in Weyl semimetals by making use of the Kubo 
linear-response theory and a simple continuum model of a semiinfinite Weyl semimetal with a pair of Weyl nodes. 
We calculated explicitly the Fermi arc quasiparticle width and found that there are contributions due to
(i) the intra-arc scattering and (ii) the bulk dephasing of the Fermi arc states [see, Eqs.~(\ref{Gamma-FA})
and (\ref{Gamma-FB}), respectively]. Furthermore, we calculated the Fermi arc conductivity explicitly and
confirmed that the scattering of the Fermi arc quasiparticles into the bulk ones is indeed responsible
for the dissipative transport. In the analysis, we used the current-current correlator with the vertex
correction in the ladder approximation. We also showed that the corresponding correction may be important
in certain regions of the parameter space. 

Our results reveal that the Fermi arc conductivity at zero temperature decreases with the growth of chemical
potential (see, the right panel of Fig.~\ref{fig:sigma-xx-y-tot}). This is explained by the associated increase of
the Fermi arc quasiparticle width that grows with the density of states in the bulk. By noting that the ladder
approximation for the vertex gradually deteriorates at small values of chemical potential $\mu$, our results
for the conductivity are quantitatively reliable only at sufficiently large $\mu$. The results at small (vanishing)
values of chemical potential $\mu$ should, therefore, be treated as an extrapolation. We hope that it would
be possible to address the corresponding regime in the future by using a non-perturbative resummation for
the vertex function. In this paper, we also analyzed the temperature dependence of the Fermi arc
conductivity, which appears to be a decreasing function of $T$ when the chemical potential is sufficiently
small (i.e., $\mu \gamma/v_F^2\lesssim 0.06$). At larger value of $\mu$, the temperature dependence of
the conductivity is nonmonotonic with a local maximum at a temperature that scales approximately as the
chemical  potential, see Fig.~\ref{fig:sigma-xx-T-tot-mu}. This suggests a certain nontrivial competition
between the phase space of the bulk states and the thermal effects. The precise nature of the corresponding
mechanism, however, remains somewhat elusive. We hope to consider this issue in more detail elsewhere.

As we showed in our study, the region of small chemical potential $\mu$ is peculiar. This is due
to the fact that the vertex function becomes large in the region of small energies and the leading-order
perturbative treatment is insufficient. In this case, a nonperturbative analysis of the integral equation
for the vertex (\ref{vertex-01}) is needed. Finally, there is an interesting question as to what happens
in the case where the chemical potential is zero. Since the Fermi arc width decreases as
chemical potential decrease, one would naively think that the Fermi arc conductivity blows
up at $\mu \to 0$. However, according to the studies performed in Refs.~\cite{Nandkishore:2014}
and \cite{Rodionov:2015}, there is still a non-zero density of the bulk states in a disordered
Weyl semimetal at zero chemical potential. In the case of charged impurities, the nonvanishing
bulk density of states is realized due to the formation of electron and hole puddles \cite{Rodionov:2015}.
The low-energy behavior of a Weyl semimetal with local impurities is dominated by rare region
effects, which are essentially nonperturbative \cite{Nandkishore:2014}. Thus, the bulk states
in disordered Weyl semimetals are present even at zero chemical potential. This suggests
that the Fermi arc states always couple to the bulk ones, and the Fermi arc conductivity is always finite.

As is clear, the model used in this paper is quite simple and may not describe perfectly the 
transport of the Fermi arcs in real materials. In order to fully appreciate its strengths and 
limitations, therefore, it is pertinent to discuss the features of the model in more detail. 
For the sake of clarity, we studied a model with only one Fermi arc. In realistic Weyl 
semimetals (e.g., $\mathrm{TaAs}$, $\mathrm{TaP}$, $\mathrm{NbAs}$, and 
$\mathrm{NbP}$) there are usually more Fermi arcs, sometimes up to 12 
\cite{Bian, Qian}. For the dissipation mechanism that we discussed, this is not 
very important, of course. However, the existence of numerous arcs opens the 
possibility for scattering processes between different Fermi arcs. Qualitatively, 
such additional processes can only increase surface quasiparticle width and, 
therefore, reduce the conductivity. It is also worth noting that the real Fermi 
arcs have nontrivial spin polarization \cite{Wang, Hasan-spin, QPI:Hasan}.
This feature is left out in our model. Whether the spin texture of Fermi arc 
states is relevant for the Fermi arc transport is probably a material dependent 
property. In this connection, we would like to note that, according to the recent 
experimental results in Ref.~\cite{QPI:Inoue}, the spin texture does play a 
qualitative, although minor, role in $\mathrm{TaAs}$. We assumed also that 
impurities are dilute so that our perturbative treatment of the self-energy and the
current-current correlator is justified. In the regime of strong disorder, one should 
use a self-consistent Born approximation, renormalization group, or other 
non-perturbative techniques. In addition, the localization effects should be taken 
into account at large disorder strength. Finally, it would be very interesting 
to generalize our study to the case of the optical conductivity of Fermi arc states. 
Note that the optical conductivity of the bulk states was recently studied in 
Refs.~\cite{Carbotte:2014,Carbotte:2016,Juricic:2016}.

\acknowledgments

The work of E.V.G. was supported partially by the Ukrainian State Foundation for Fundamental
Research. The work of V.A.M. was supported by the Natural Sciences and Engineering
Research Council of Canada. The work of I.A.S. was supported in part by the
U.S. National Science Foundation under Grant No.~PHY-1404232.

\end{document}